\documentclass[longauth,traditabstract]{aa}

\usepackage{graphicx}
\usepackage{txfonts}
\usepackage{comment}
\usepackage{lscape}
\usepackage{longtable}

\usepackage{natbib,twoopt}
\bibpunct{(}{)}{;}{a}{}{,}             

\def\ojo{\fbox{\bf !$\odot$j$\odot$!}} 			

\usepackage[usenames]{color}                                   
\def\ATT{\color{red}}                				
\def\ATC{\color{green}}                				
\def\ATC{\color{cyan}}                				

\newcommand\pycasso{{\sc p}y{\sc casso}}          	
\newcommand\starlight{{\sc starlight}}          	

\begin{document}

\title{Star formation along the Hubble sequence:} 
\subtitle{Radial structure of the star formation of CALIFA galaxies}

\authorrunning{IAA et al.}
\titlerunning{Star formation rate}

\author{
R. M. Gonz\'alez Delgado\inst{1},
R. Cid Fernandes\inst{2},
E. P\'erez\inst{1},
R. Garc\'{\i}a-Benito\inst{1}, 
R. L\'opez Fern\'andez\inst{1},
E. A. D. Lacerda\inst{1,2},
C. Cortijo-Ferrero\inst{1},
A. L.\ de Amorim\inst{2},
N. Vale Asari\inst{2},
S. F. S\'anchez\inst{3},
C. J. Walcher\inst{4},
L. Wisotzki\inst{4},
D. Mast\inst{5},
J. Alves\inst{6},
Y. Ascasibar\inst{7, 8},
J. Bland-Hawthorn\inst{9},
L. Galbany\inst{10,11},
R. C. Kennicutt Jr.\inst{12},
I. M\'arquez\inst{1},
J. Masegosa\inst{1},
M. Moll\'a\inst{13},
P.  S\'anchez-Bl\'azquez\inst{7, 8},
J. M. V\'{\i}lchez\inst{1},
CALIFA collaboration\inst{14}
}

\institute{
Instituto de Astrof\'{\i}sica de Andaluc\'{\i}a (CSIC), P.O. Box 3004, 18080 Granada, Spain. (\email{rosa@iaa.es})
\and
Departamento de F\'{\i}sica, Universidade Federal de Santa Catarina, P.O. Box 476, 88040-900, Florian\'opolis, SC, Brazil
\and
Instituto de Astronom\'\i a,Universidad Nacional Auton\'oma de M\'exico, A.P. 70-264, 04510 M\'exico D.F., Mexico
\and
Leibniz-Institut f\"ur Astrophysik Potsdam (AIP), An der Sternwarte 16, D-14482 Potsdam, Germany
\and
Instituto de Cosmologia, Relatividade e Astrof\'{i}sica - ICRA, Centro Brasileiro de Pesquisas F\'{i}sicas, Rua Dr.Xavier Sigaud 150, CEP 22290-180, Rio de Janeiro, RJ, Brazil
\and
University of Vienna, T\"urkenschanzstrasse 17, 1180, Vienna, Austria
\and
Depto. de F\'{\i}sica Te\'orica, Universidad Aut\'onoma de Madrid, 28049 Madrid, Spain
\and 
CAB: "Astro-UAM, UAM, Unidad Asociada CSIC", 28049 Madrid, Spain
\and
Sydney Institute for Astronomy, The University of Sydney, NSW 2006, Australia
\and
Millennium Institute of Astrophysics and Departamento de Astronom\'\i a, Universidad de Chile, Casilla 36-D, Santiago, Chile
\and
Departamento de Astronom\'ia, Universidad de Chile, Casilla 36-D, Santiago, Chile
\and
University of Cambridge, Institute of Astronomy,
Madingley Road, Cambridge, CB3 0HA, United Kingdom
\and
Departamento de Investigaci\'on B\'asica, CIEMAT, Avda.\ Complutense 40, E-28040 Madrid, Spain
\and
http://califa.caha.es
}

\date{Jan 2016}


\abstract{
The spatially resolved stellar population content of today's galaxies holds important information to understand the different processes that contribute to the star formation and mass assembly histories of galaxies.
The aim of this paper is to characterize the radial structure of the star formation rate (SFR) in galaxies in the nearby Universe as represented by a uniquely rich and diverse data set drawn from the CALIFA survey.
The sample under study contains 416 galaxies observed with integral field spectroscopy, covering a wide range of Hubble types and stellar masses ranging from $M_\star \sim 10^9$ to $7 \times 10^{11} M_\odot$. 
Spectral synthesis techniques are applied to the datacubes to derive 2D maps and radial profiles of the intensity of the star formation rate in the recent past ($\Sigma_{\rm SFR}$), as well as related properties such as the local specific star formation rate (sSFR), defined as the ratio between $\Sigma_{\rm SFR}$ and the stellar mass surface density ($\mu_\star$).
To emphasize the behavior of these properties for galaxies that are on and off  the main sequence of star formation (MSSF) we stack the individual radial profiles in seven bins of galaxy morphology (E, S0, Sa, Sb, Sbc, Sc, and Sd), and several stellar masses. 
Our main results are: 
{\em (a)} The intensity of the star formation rate shows declining profiles that exhibit very little differences between spirals, with values at $R = 1$ half light radius (HLR) within a factor two of $\Sigma_{\rm SFR} \sim 20\, M_\odot\,$Gyr$^{-1}\,$pc$^{-2}$.
The dispersion in the $\Sigma_{\rm SFR}(R)$ profiles is significantly smaller in late type spirals (Sbc, Sc, Sd).  This  confirms that the MSSF is a sequence of galaxies with nearly constant $\Sigma_{\rm SFR}$.
{\em (b)} sSFR values scale with Hubble type and increase radially outwards, with a steeper slope in the inner 1 HLR. This behavior suggests that galaxies are quenched inside-out, and that this process is faster in the central, bulge-dominated  part than in the disks. 
{\em (c)} As a whole, and at all radii, E and S0 are off the MSSF, with SFR much smaller than spirals of the same mass.
{\em (d)} Applying the volume-corrections for the CALIFA sample, we obtain a density of star formation in the local Universe of $\rho_{\rm SFR} = (0.0105 \pm 0.0008) \, M_\odot\,$yr$^{-1}\,$Mpc$^{-3}$, in agreement with independent estimates. 
Most of the star formation is occurring in the disks of spirals.
{\em (e)} The volume averaged birthrate parameter, that measures the current SFR with respect to its lifetime average,
$b^\prime = 0.39 \pm 0.03$, suggests that the present day Universe is forming stars at $\sim 1/3$ of its past average rate. E, S0, and the bulge of early type spirals (Sa, Sb) contribute little to the recent SFR of the Universe, which is dominated by the disks of Sbc, Sc, and Sd spirals.
{\em (f)} There is a tight relation between $\Sigma_{\rm SFR}$ and $\mu_\star$, defining a local MSSF relation with a logarithmic slope of 0.8, similar to the global MSSF relation between SFR and $M_\star$. This suggests that local processes are important in determining the star formation in disks, probably through a density dependence of the SFR law. 
The scatter in the local MSSF  is driven  by morphology-related offsets, with $\Sigma_{\rm SFR} / \mu_\star$ (the local sSFR) increasing  from early to late type galaxies, indicating that the shut down of the star formation is more related with global processes, such as the formation of a spheroidal component. 
}



\maketitle

\section{Introduction}
\label{sec:Introduction}

Nearly a century later, the  simple classification scheme introduced by \citet{hubble36}  is still in use. The reason it remains useful is that the physical properties of galaxies correlate with the morphology in a broad context \citep[e.g.][]{holmberg58, roberts63, roberts94}. In particular, the Hubble sequence can be described as a sequence in terms of recent star formation, increasing in importance from elliptical (E) to spiral (S) galaxies \citep[e.g.][]{kennicutt83, kennicutt98}.

Contemporary galaxy surveys have mapped this bimodal distribution implicit in the Hubble classification in terms of properties related with their structure (morphology), stellar content, and chemical composition \citep[e.g.][]{blanton03, baldry04, blanton09, kauffmann03a, mateus06, ascasibar11, casado15}. 
One population, located in the region of the color-magnitude diagram (CMD) known as the red sequence, is composed of galaxies with little star formation, large stellar masses ($M_\star$), stellar mass surface density ($\mu_\star$), and light concentration. The other major population,
located in the so called blue cloud in the CMD, consists of galaxies with significant star formation, smaller $M_\star$ and $\mu_\star$,  and small concentration in light. The red sequence is the realm of early type galaxies, whereas galaxies of later Hubble types populate the blue cloud. 

Other works have shown that galaxies that populate the blue cloud follow a  strong correlation between $M_\star$ and the present star formation rate (SFR) \citep{brinchmann04, salim07, renzinipeng15, catalan15}; the main-sequence of star forming galaxies (MSSF).
The correlation is tight, with only 0.2--0.3 dex dispersion in SFR for a fixed $M_\star$, and with a slope that is somewhat smaller than unity, implying  that the relative rate at which stars form in galaxies, i.e. the specific star formation rate sSFR = SFR/$M_\star$, declines weakly with increasing galaxy mass \citep{salim07, schiminovich07}.
 
Subsequent studies have shown that the MSSF relation persists to at least $z \sim 4$ \citep[e.g.][]{noeske07, daddi07, elbaz07, peng10, wuyts11}. These works conclude that most of the star formation in the Universe is  produced by galaxies in the main sequence, with starbursts (which deviate upwards from the MSSF) contributing with only $\sim 10 \%$ to the total star formation rate at $z \sim 2$ \citep{sandersmirabel96,  rodighiero11}, where the peak of the cosmic star formation rate occurs \citep[e.g.][]{madau14}. 
A recent study by \citet{speagle14} finds that the logarithmic slope of the MSSF relation increases with cosmic time, from $\sim 0.6$ at $z \sim 2$ to $0.84$ at $z=0$. This implies that the characteristic sSFR of the main sequence population evolves rapidly with redshift \citep{karim11}. In fact, \citet{elbaz11} show that star formation has decreased by a factor of 20 since $z \sim 2$, and that the corresponding sSFR declined as $t^{-2.2}$, where $t$ is the cosmic epoch. 

There is also a substantial population of quenched galaxies that dominate the high end of the mass function, but whose sSFR is significantly lower than in star forming galaxies \citep{salim07, schiminovich07, chang15}.  
In a simple picture, galaxies evolve along the blue star forming MSSF, increasing in mass through the accretion of cold gas from the cosmic web and/or through mergers. When it approaches a critical mass the supply of  gas is shut off. Star formation is thus quenched and the galaxy migrates to the red sequence, where the increase in mass and size may happen through minor mergers \citep[e.g.][]{faber07, lilly13}. Although the quenching phase 
is relevant in the life of a galaxy, it is not clear at which critical mass the galaxy is quenched, and whether this is related with a change in the nature of the gas accretion, with heating of the surrounding gas by an AGN, or with the formation of an spheroidal component \citep{martig09}.

Evidently, this whole field relies on empirical measures of the SFR. There is no shortage of methods to estimate the SFR, each with its virtues and caveats (see \citet{kennicuttevans12} for a review).  Some gauge the SFR indirectly by quantifying how the radiative output of young stars is reprocessed by gas or dust, as H$\alpha$ and far infra-red SFR indicators. Direct detection of recently formed stars is best done in the UV, where they outshine older populations by large factors, although dust inevitably introduces uncertainties.
Because of the comparable contributions of stellar generations of all ages, the optical continuum is not the cleanest spectral range to work with if one is interested in pinning down the recent star formation history (SFH). 
It is, nonetheless,  the very spectral range where galaxy evolution first started and matured as a research field, as illustrated by the seminal works of \citet{tinsley68, tinsley72}, \citet{searle73},  \citet{gallagher84}, and \citet{sandage86}, who first used  galaxy optical colors to study how SFHs vary along the Hubble sequence and to predict the cosmic evolution of the SFR. 
This line of work has been revamped in the last decade or two with the confluence of advances in the spectral modeling of stellar populations \citep[e.g.][]{starbursts99, bruzual03, gonzalezdelgado05, maraston05, vazdekis10}, the development of full spectral synthesis methods \citep[e.g.][]{panter03, cidfernandes05, ocvirk06, koleva11, sanchez15b}, and the flood of data from surveys such as the SDSS \citep{sdss}, 
which provided abundant observational material to explore these new tools \citep[e.g.][]{panter03, heavens04, asari07, panter08,  tojeiro11}.
A detailed discussion of the uncertainties associated to these methods can be found in the recent reviews by \citet{walcher11} and \citet{conroy13}.

Regardless of the method employed to derive SFHs and SFRs, an important limitation of most studies to date is the lack of spatially resolved information. Galaxies are usually studied as a whole, with observations integrated over their distinct morphological components, or else with data that only partially cover them and are thus prone to aperture effects.
Overcoming this limitation requires data of the kind that only recently started to become available with Integral Field Spectroscopy (IFS)  surveys such as ATLAS3D \citep{cappellari11}, CALIFA \citep{sanchez12, husemann13, garciabenito15}, SAMI \citep{bryant15}, and MaNGA \citep{bundy15}. These new generation surveys are a step forward to understand the star formation in galaxies, and should help us disentangling the contributions of spheroids and disks to the MSSF relation.

Because of its focus on early type galaxies (E, S0, Sa), ATLAS3D essentially avoids star forming systems, and hence does not constitute an ideal sample to study the MSSF \citep{mcdermid15}. CALIFA, on the other hand, is particularly well suited for this study.
Firstly, in includes a large, homogeneous but diverse sample of galaxies covering the full Hubble sequence, from Ellipticals (E0-E7), and Lenticulars (S0-S0a), to Spirals (Sa to Sd), and a correspondingly large range of masses ($10^9$ to $\sim 10^{12}  M_\odot$, \citealt{gonzalezdelgado14a}). 
Second, its large field of view ($74{\tt''} \times 64{\tt''}$,  with final spatial sampling of $1{\tt''}$) covers the full extent of the galaxies and allows us to spatially map the star formation, as well as to obtain the total integrated SFR. 
Third, it covers the whole rest-frame optical wavelength at intermediate spectral resolution, that allow us to apply full spectral fits to retrieve the SFHs (and thus recent SFR too).
Finally, the volume-corrected distribution functions of CALIFA are fully compatible with estimates from the full SDSS when accounting for large-scale structure \citep{walcher14}, which allows the extrapolation of results to the overall cosmic context.

Previous papers in this series  used the SFHs of  $\sim100$--300 galaxies of the CALIFA survey to derive spatially resolved information on the mass growth of galaxies \citep{perez13}, and stellar population properties like the stellar mass surface density, ages, stellar metallicity, and extinction \citep{gonzalezdelgado14a, gonzalezdelgado14b, gonzalezdelgado15}. 
We found that massive galaxies grow their stellar mass inside-out, where the signal of downsizing is spatially preserved, with both inner and outer regions growing faster for more massive galaxies. We confirm that more massive galaxies are more compact, older, more metal rich, and less reddened by dust. Additionally, we find that these trends are preserved spatially with the radial distance to the nucleus. Deviations from these relations appear correlated with Hubble type: earlier types are more compact, older, and more metal rich for a given $M_\star$, an indication that quenching is related to morphology.

Here we concentrate on the study of the ongoing star formation of CALIFA galaxies, as derived from full spectral fits of the optical stellar continuum. The goals are: 1)  To characterize in detail the radial structure of the SFR and sSFR of galaxies in the local Universe; 2) to examine how SFR and sSFR relate to Hubble type; 3) to spatially resolve the MSSF relation; 4) to estimate the contribution of different types of galaxies and their sub-components to the cosmic star formation rate.

This paper is organized as follows: Section \ref{sec:Data} describes the observations and  summarizes the properties of the galaxies analyzed here. In Section \ref{sec:Method} we summarize our method for extracting the SFH and explain how we measure the present SFR. Section \ref{sec:GlobalMSSF} presents results on the MSSF relation and how our assumptions affect it. Section \ref{sec:RadialStructures} deals with the radial structure of the intensity of the star formation rate ($\Sigma_{\rm SFR}$), and related properties such the local specific SFR. We discuss the results and their relation with the cosmic star formation of the local Universe in Section \ref{sec:Discussion}. Section \ref{sec:Summary}  summarizes our main findings.

\section{Data and sample}
\label{sec:Data}

\subsection{Observations and data reduction}
\label{sec:Observations}

The observations were carried out with the Potsdam Multi-Aperture Spectrometer PMAS \citep{Roth05} in the PPaK mode \citep{verheijen04} at the 3.5m telescope of Calar Alto observatory. PPaK contains 382 fibers of $2.7^{\prime\prime}$  diameter each, and a  $74^{\prime\prime} \times 64^{\prime\prime}$ 
FoV  \citep{kelz06}. Each galaxy is observed with two spectral settings, V500 and V1200, with spectral resolutions  $\sim 6$ (FWHM) and 2.3 \AA, respectively. The V500 grating covers from 3745 to 7300 \AA, while the V1200 covers 3650--4840 \AA. 
To reduce the effects of vignetting on the data, we combine the observations in the V1200 and V500 setups,  calibrated with version 1.5 of the reduction pipeline. We refer to \citet{sanchez12}, \citet{husemann13}, and \citet{garciabenito15} for  details on the observational strategy and data processing.

\subsection{Sample: morphological classification}
\label{sec:Sample}

The CALIFA mother sample consists of 939 galaxies selected from the SDSS survey in the redshift range $z = 0.005$--0.03, and with $r$-band angular isophotal diameter of 45--80$^{\prime\prime}$. It is primarily a diameter-limited sample to guarantee that the objects fill the $74^{\prime\prime} \times 64^{\prime\prime}$ FoV. It includes a significant number of galaxies in different bins in the CMD, and covers a wide and representative range of galaxy types. 
The galaxies were morphologically classified by five members of the collaboration through visual inspection of the SDSS $r$-band images, averaging the results (after clipping outliers). The sample and its characteristics are fully described in \citet{walcher14}. 

The targets studied in this paper were selected from those observed in both V1200 and V500 setups earlier than January 2015, and excluding type 1 Seyferts and galaxies that show merger or interaction features. This leaves a final sample of  416 galaxies.

\begin{figure}
\includegraphics[width=0.48\textwidth]{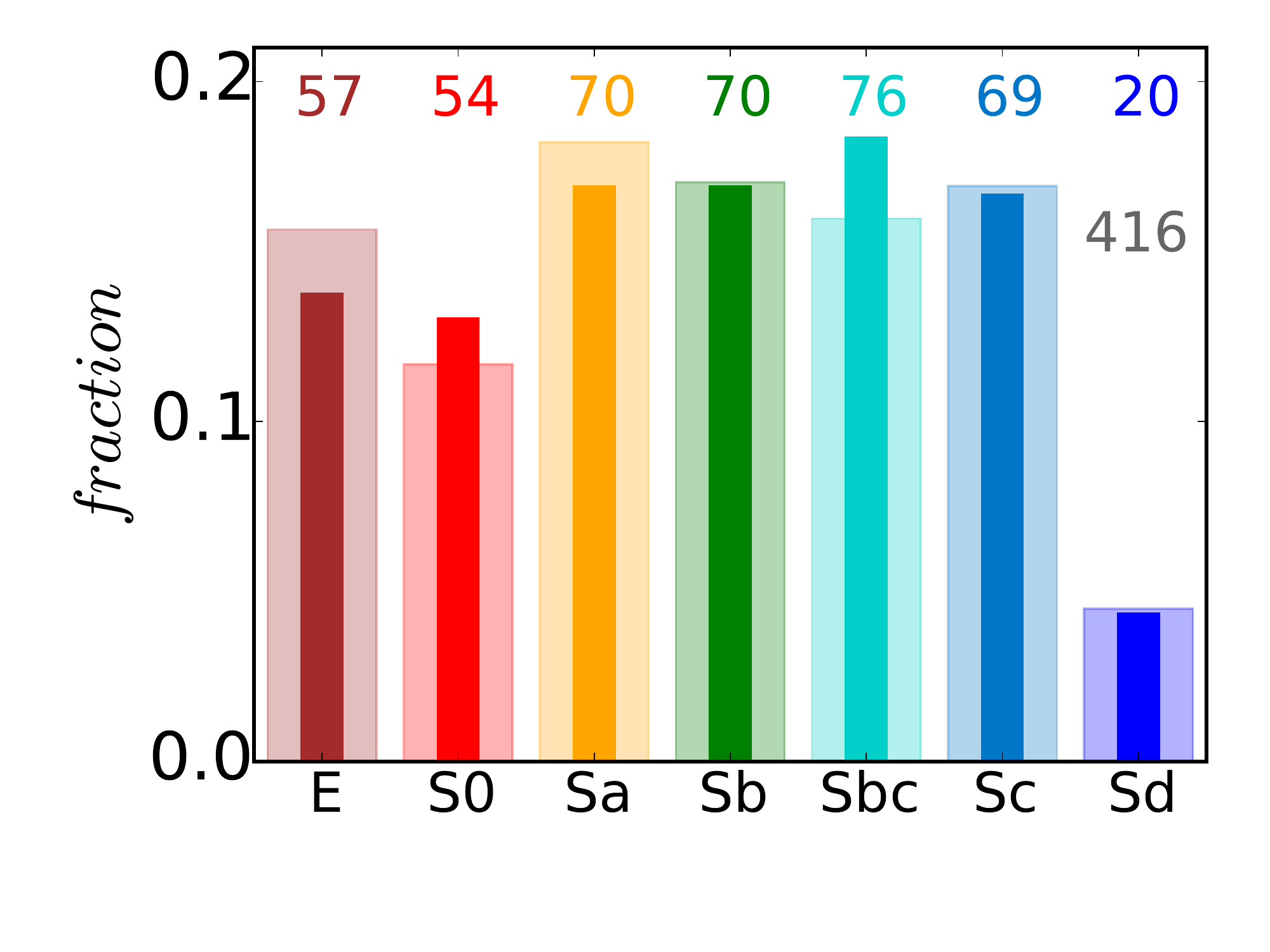}
\caption{Comparison of the distribution of Hubble types in the CALIFA mother sample (939 galaxies,  bars) and the 416 galaxies analyzed here (filled narrow color bars). 
The histograms are normalized to unity, so that the two distributions are directly comparable.
The number of galaxies in each morphology bin is labeled in color, with the same palette used throughout the paper.
 }
\label{fig:hist-type}
\end{figure}

As we have done in GD15, we group the galaxies into seven morphology bins: E (57 galaxies), S0 (54, including S0 and S0a), Sa (70, including Sa and Sab), Sb (70), Sbc (76), Sc (69, including Sc and Scd), and Sd (20, including 18 Sd, 1 Sm, and 1 Irr). Fig.\ \ref{fig:hist-type} shows the morphological distribution of our 416 galaxies (filled bars) as well as that of the mother sample (empty bars). 
Hubble types are labeled with a brown to blue (ellipticals to late type spirals) color palette, used throughout this paper.
The similarity of the two distributions ensures that our sub-sample is a fair representation of the mother sample. This is an important aspect, as it allows us to apply the volume corrections derived by \citet{walcher14} to extend the statistical results presented here to those of the galaxy population as a whole.

\section{Stellar population analysis: Mass and star formation rate}
\label{sec:Method}

\subsection{Method of analysis}
\label{sec:Base}

To extract the stellar population properties from the datacubes we apply the same method than in  \citet{perez13}, \citet{cidfernandes13,cidfernandes14}, and \citet{gonzalezdelgado14a,gonzalezdelgado14b, gonzalezdelgado15}. 
Briefly, after some basic pre-processing steps like spatial masking of foreground and background sources, rest-framing,  and spectral resampling, the individual spectra that have signal-to-noise ratio ${\rm S/N} \leq 20$ in a 90 \AA\ window centered at 5635 \AA\ (rest-frame), are coadded into Voronoi zones \citep{cappellari03}.  The resulting 366112 spectra (880 per galaxy, on average) are then fitted with \starlight\ \citep{cidfernandes05}
using the cluster Grid-CSIC at the Instituto de Astrof\'\i sica de Andaluc\'\i a. The output is then processed through \pycasso\ (the Python CALIFA \starlight\ Synthesis Organizer) to produce a suite of the spatially resolved stellar population properties.

The base used in \starlight's spectral decomposition is a central ingredient in our whole analysis. The results presented here were obtained with 
base {\em GMe}, as defined in \citet{gonzalezdelgado14a,gonzalezdelgado14b,gonzalezdelgado15}. This base comprises 235 spectra for simple stellar populations (SSP) drawn from \citet{vazdekis10} for populations older than $t = 63$ Myr and from \citet{gonzalezdelgado05} models for younger ages. The evolutionary tracks are those of \citet{girardi00}, except for the youngest ages (1 and 3 Myr), which are based on the Geneva tracks \citep{schaller92,schaerer93,charbonnel93}. The Initial Mass Function (IMF) is Salpeter. The $Z$ range covers the seven metallicities provided by \citet{vazdekis10} models: $\log Z/Z_\odot = -2.3$, $-1.7$, $-1.3$, $-0.7$, $-0.4$, 0, and $+0.22$, but SSPs younger than 63 Myr include only the four largest metallicities. The Appendix presents some comparisons with results obtained with an alternative base built from a preliminary update of the \citet{bruzual03} models.

\subsection{Stellar masses}

Our galaxy stellar masses ($M_\star$) are obtained by adding the masses of each spatial zone \citep{cidfernandes13,gonzalezdelgado15}. This procedure takes into account spatial variations of the  stellar population properties and stellar extinction, something which cannot be done when dealing with integrated light data (i.e., one spectrum per galaxy). Masked spaxels, due to foreground stars or other artifacts, are corrected for in \pycasso\ using the stellar mass surface density ($\mu_\star$) radial profile as explained in \citet{gonzalezdelgado14a}. 

Both $M_\star$ and Hubble type play important roles in this paper, so it is important to know how these two properties relate to each other.
Table \ref{tab:Massdistribution} shows the distribution of galaxies by Hubble type in several bins of $M_\star$. 
The masses range from $8 \times10^8$ to $7 \times 10^{11} M_\odot$ (for a Salpeter IMF), and peak at $\sim 10^{11} M_\odot$. 
As expected, $M_\star$ correlates with Hubble type (see also \citealt{gonzalezdelgado15}, particularly their Fig.\ 2). 
E are the most massive galaxies with $\log M_\star = 11.3  \pm 0.3$ (average $\pm$ dispersion) in solar units,  and the least massive galaxies are those in the Sd bin, with  $\log M_\star = 9.6 \pm 0.4$. The more typical CALIFA galaxy has $\log M_\star  = 10.75$,  similar to the Milky Way's mass \citep{licquia15}.

\label{sec:Mass_x_Morphology}
\begin{table}
\caption{Number of galaxies in each Hubble type and mass interval.}
\begin{tabular}{lccccccc}
\hline\hline
$\log  M_\star (M_\odot)$ &   E & S0 & Sa & Sb & Sbc & Sc & Sd    \\      
\hline
$\leq 9.1$   & -  & -     & -     & -    & -     &    1   & 1 \\
~~9.1--9.6       & -  &      & -     & -    & -      & 15 & 11 \\
~~9.6--10.1     & -  &  1     &  2   & -    &  5  & 14 & 5 \\
10.1--10.6   & 1  &  3   & 10  & 14  & 18 & 28 & 3 \\
10.6--10.9   & 7  &  15 & 10  & 19  & 29 &  4  & 0 \\
10.9--11.2   & 16 & 18 & 28  &  22 & 20 &  6  & - \\
11.2--11.5   & 19 & 15 & 20  &  14 &  4  &  1  & - \\
11.5--11.8   & 13 &  2  &  -    &    1 & -    &  -   & - \\
$\geq11.8$ & 1   & -    &  -    & -     & -    & -   & - \\
\hline
 total (416)  & 57 & 54 & 70   & 70  & 76 & 69 & 20 \\
 \hline
 $\langle \log M_\star \rangle$ & 11.3 & 11.0 & 11.0 & 10.9 & 10.7 & 10.1 & 9.6\\
 $\sigma ( \log M_\star)$ & 0.3 & 0.3 & 0.4 & 0.3 & 0.3 & 0.5 & 0.4 \\
\hline\hline
\label{tab:Massdistribution}
\end{tabular}
\end{table}

\subsection{Estimation of the recent SFR from the spectral synthesis }

\begin{figure}
\includegraphics[width=0.48\textwidth]{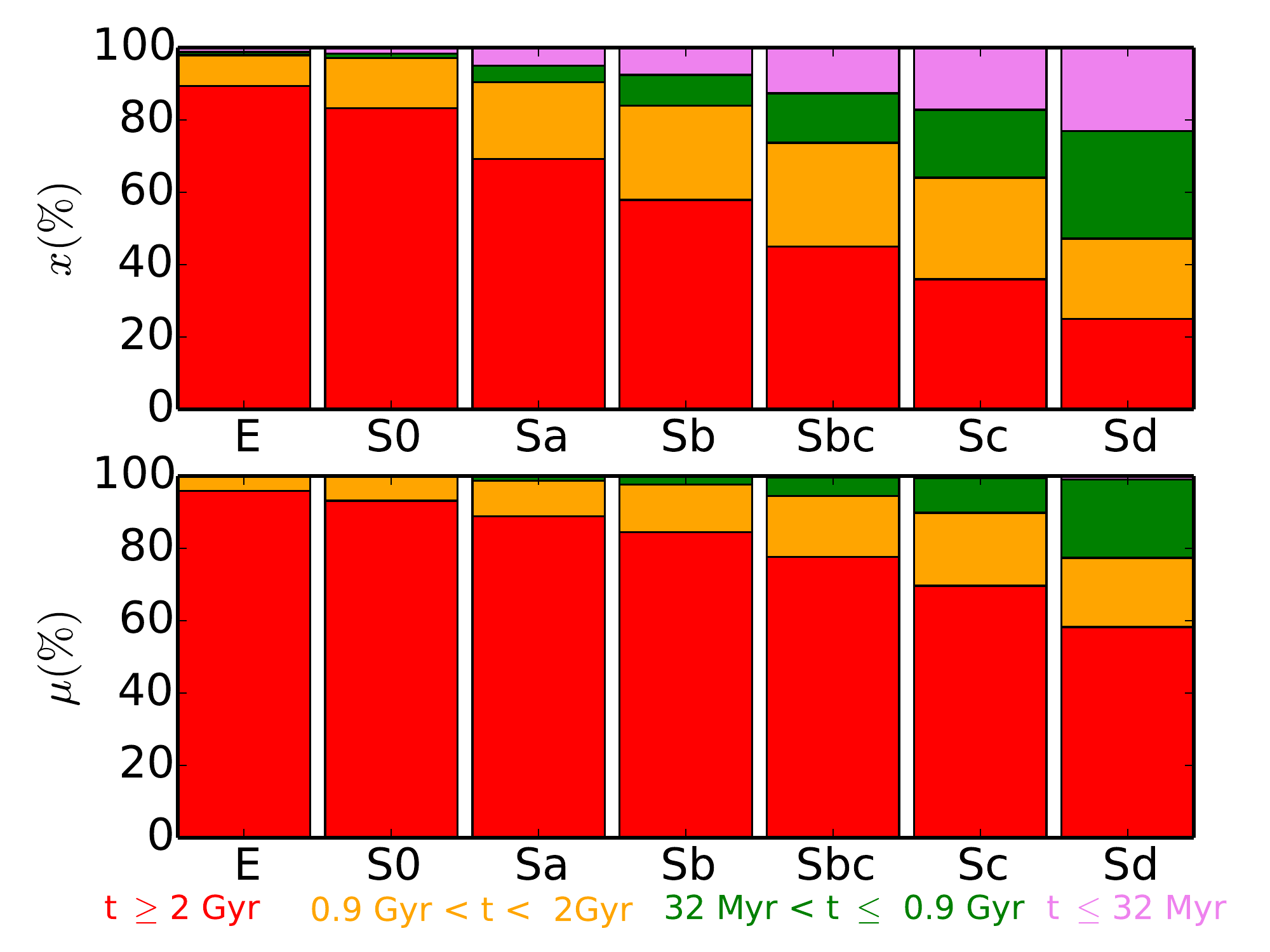}
\caption{Average light (upper panel) and mass (bottom) fractions (defined with respect to $\lambda$ = 5635 \AA, the normalizing wavelength) due to stars in different age ranges as a function of  Hubble type.
Age ranges are coded by color: The youngest ones, $< 32$ Myr, in violet (hardly visible in the bottom panel because they carry little mass). Populations from 32 to 900 Myr are shown in green; those from 0.9 to 2 Gyr in orange, and older ones in red.}
\label{fig:histpopx}
\end{figure}

SFR is usually estimated from H$\alpha$, far-infrared, or UV luminosities \citep{kennicutt98, kennicuttevans12, catalan15}, which, despite their own caveats and limitations, get the job done with conveniently simple, one-line formulae. No such straightforward recipe exists for optical continuum data, however. The reason is that stars of all ages can make comparable contributions to the optical light, and isolating the part due to those formed in the recent past is not a trivial task. It is, however, a feasible one. After all, decomposing a spectrum in terms of stellar populations of different ages is precisely what \starlight\ does. 
In fact, an extended version of the code is being developed which incorporates UV, far-infrared, and/or emission line information (L\'opez Fern\'andez et al.\ 2016), all of which should improve its sensitivity to young stars. In any case, as shown by  \citet{asari07}, the standard version of  \starlight\ already performs well in this respect.

This section explains our methodology to compute SFRs. The SFR values themselves are presented in later sections, while the discussion here focuses on how to handle the  \starlight\  output to produce meaningful SFR estimates, the uncertainties involved, and how to improve the results by means of criteria based on ancillary emission line information.

\subsubsection{Choice of a ''recent" star formation time scale}

Let us first specify what we mean by ``recent past'' by defining $t_{SF}$ as the age of the oldest stars to be included in the computation of our recent SFR. The mean rate of star formation then follows from a simple summation over all populations younger that $t_{SF}$:

\begin{equation} 
\label{eq:SFR}
{\rm SFR}_{xy} =  \frac{1}{t_{SF}}  \sum_{t \leq t_{SF}} M_{txy} 
\end{equation}

\noindent where $xy$ denotes a spaxel (or Voronoi zone) and $M_{txy}$ is the mass initially turned into stars which now have age $t$ at the same $xy$ location. Radial profiles of SFR and galaxy wide rates are trivially obtained by averaging ${\rm SFR}_{xy}$ over the desired $xy$ region. Similarly, surface densities ($\Sigma_{\rm SFR}$) are obtained by dividing by the corresponding area (a spaxel, a radial ring, the whole galaxy, etc.).

The choice of $t_{SF}$ is arbitrary, so let us sketch some general guidelines to choose a useful value. Naturally, the larger $t_{SF}$ is the more robust the corresponding SFR becomes, since more base elements are summed over in eq.\ \ref{eq:SFR}, thus minimizing known degeneracies in stellar population synthesis (e.g., \citealt{cidfernandes14}). On the other hand, one would like $t_{SF}$ to be much smaller than the Hubble time (otherwise SFR and $M_\star$ become $\sim$ equivalent quantities). Furthermore, it would be desirable to have a $t_{SF}$
 of the same order of the time scale involved in some other independent SFR tracer to which we can compare ours. The natural choice of reference in our case is H$\alpha$, first because of the wide spread use of this tracer, but also because we lack UV or far infrared data with CALIFA-like spatial resolution for our sample. The H$\alpha$ luminosity responds to the $h\nu > 13.6$ eV radiation field, which is completely dominated by $\la 10$ Myr populations, so we aim at  $t_{SF}$ of this same order of magnitude.

After some experimentation we chose $t_{SF} = 32$ Myr\footnote{
This overly precise looking value of $t_{SF}$ merely reflects the choice of which ages in our discrete grid to include in the summation in eq.\ (\ref{eq:SFR}). We chose to include up to the base element at $t = 32$ Myr  (actually 31.62 Myr).  
}.
This choice follows the same rationale  (but different data) as in \citet{asari07}, who,
in a study of star-forming galaxies in the SDSS,  found that a very similar time-scale (25 Myr; see their figure 6) produces the best correlation between \starlight\ and H$\alpha$-based estimates of SFR. We defer a detailed discussion of this point to a future communication (Lacerda et al., in prep.). For the purposes of this paper it suffices to say that  $t_{SF}$-values between $\sim 10$ and 100 Myr would lead to the same overall qualitative conclusions.

\subsubsection{SFHs along the Hubble sequence: a condensed view}

Fig.\ \ref{fig:histpopx} tracks the percent contribution in light (top panels) and mass (bottom) of our recent populations ($\leq t_{SF} = 32$ Myr, in magenta) along the Hubble sequence, as well as those of stars in three other intervals: 32 Myr to 0.9 Gyr (green), 0.9 to 2 Gyr (orange), and $\geq$ 2 Gyr (red). 
These four intervals roughly represent populations in which the light is dominated by O, B, A--early F, and later type (lower mass) stars.
This strategy of grouping stellar populations in broad age-ranges as a way of summarizing SFHs goes back to early  studies based on equivalent widths and colors \citep{bica88, bica94, cidfernandes01}, but was also applied in full spectral fitting work \citep{gonzalezdelgado04, cidfernandes04, cidfernandes05}. 

The top panels in Fig.\ \ref{fig:histpopx} show a steady progression of young and intermediate age populations along the Hubble sequence. The percent light contribution (at $\lambda = 5635$ \AA) of  populations younger than 32 Myr decreases from $x_{Y} (\%) = 23.0$ in Sd galaxies to 17.2 in Sc, 12.6 in Sbc, 5.5 in Sb, and 4.9 in Sa. Because of their low mass-to-light ratio as well as the tiny time-span compared to other bins,
these populations are essentially invisible in the bottom panels, where mass fractions are plotted. Indeed, in terms of mass fractions the $< 32$ Myr populations account for only 0.97, 0.59, 0.36, 0.12, and 0.05\% for Sd, Sc, Sbc, Sb, and Sa, respectively.

\subsubsection{Star formation in early type galaxies}

The contribution of young stars decreases even further towards E and S0, but it is not null, with $x_Y \sim 2\%$. 
Naturally, the reality of populations accounting for so little light is questionable. 
Based on the extensive set of simulations carried out by \citet{cidfernandes14} to evaluate uncertainties in the \starlight\ results for CALIFA-like data, we estimate the level of noise-induced uncertainty in $x_{Y}$ to be of the order of 3\%.
Given this, the small $x_Y$ fractions identified in E and S0 should be considered noise. 
However, this 3\% error estimate reflects the level of uncertainty expected for a single spectral fit, whereas the $x_Y \sim 2\%$ in the top left of Fig.\ \ref{fig:histpopx} reflects an average over 115927 zones inside of the central 3 HLR of 111 E and S0 galaxies.
Looking from this statistical angle one should perhaps take the small $x_Y$ fractions in these systems as a sign that they may not be so quiescent after all.
There is in fact evidence for some level of star formation in at least some early type galaxies \citep{kaviraj07a}. In the context of CALIFA data, \citet{gomes15a} have unveiled spiral-arm like features consistent with recent star formation in three early type galaxies.

In any case, at such low $x_Y$-levels one also needs to worry about systematic effects, and the study by \citet{ocvirk10} is specially relevant in this respect. He finds that blue horizontal branch stars can easily masquerade as massive young stars in spectral fits, creating the artificial impression of recent star formation in otherwise genuinely old populations. This same effect was in fact detected in previous \starlight-based work on both globular clusters \citep{cidfernandes10} and passive galaxies  \citep{cidfernandes11}, and ultimately reflects limitations in the modeling of stellar evolution embedded in the SSP models used in our spectral decomposition.

As will soon become clear, for the purposes of this paper the exact values of SFR or $\Sigma_{\rm SFR}$ in E and S0 galaxies are not as important as the fact that their star forming properties are markedly different from those of later type galaxies, a relative behavior which is safely immune to the uncertainties discussed above.

\subsubsection{The equivalent width of H$\alpha$ as an ancillary constraint}

\begin{figure*}
\includegraphics[width=\textwidth]{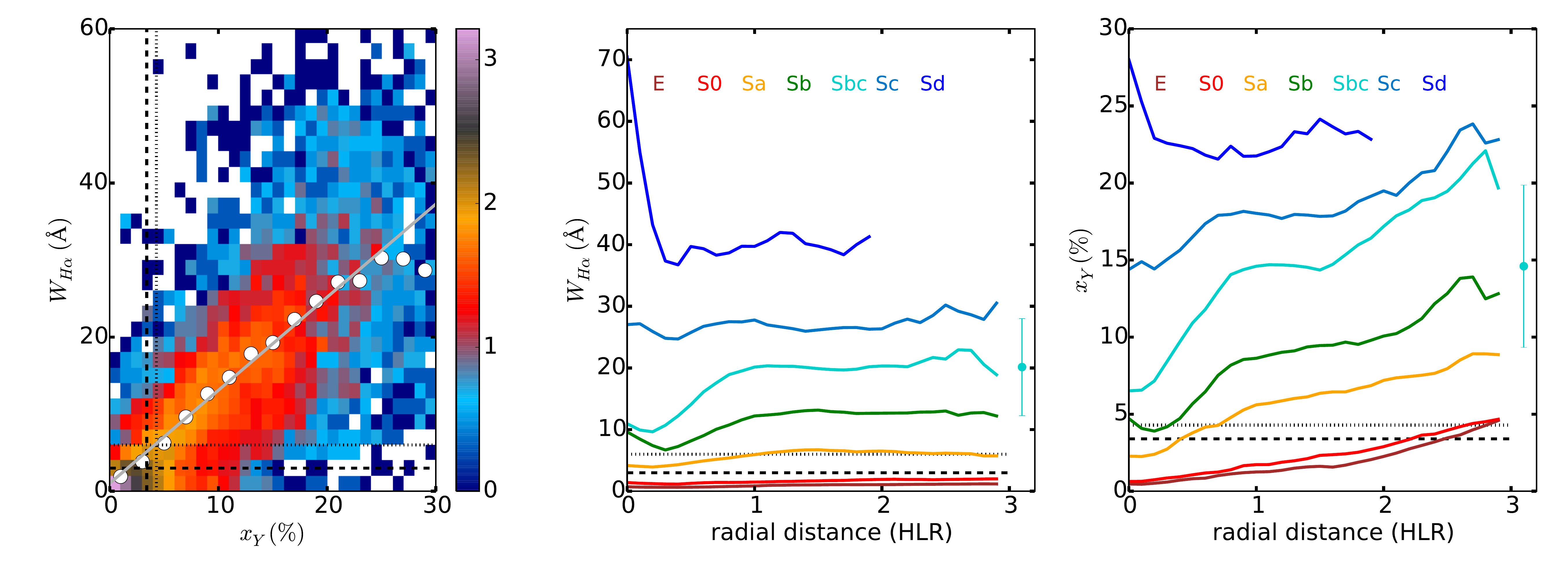}
\caption{
{\em Left:} Point density diagram of the equivalent width of the H$\alpha$ nebular emission, $W_{H\alpha}$, versus the light fraction due to populations younger than 32 Myr ($x_{Y}$) for 12540 radial bins of 416 galaxies. 
The color bar shows the density of data points in logarithm scale. White circles trace the mean relation (obtained from the mean $W_{H\alpha}$ values in $x_{Y}$-bins). The grey line shows the linear fit to all points: $W_{H\alpha}({\rm \AA}) = 1.024 + 1.208 \ x_{Y}(\%)$.  Horizontal lines are drawn at $W_{H\alpha}$ = 3 (dashed) and  6 \AA\ (dotted). Vertical lines mark $x_{Y} = 3.4\%$ (the mean $x_{Y}$ for points with $W_{H\alpha} < 3$ \AA; dashed) and
$x_{Y} = 4.3\%$ (the mean $x_{Y}$ for points with $W_{H\alpha} < 6$ \AA; dotted).
{\em Middle:} Mean $W_{H\alpha}$ radial profiles for galaxies in the seven Hubble type bins.
Horizontal lines mark $W_{H\alpha} = 3$ (dashed), and  6 \AA\ (dotted). {\em Right:} Radial profile of  $x_{Y}$ in the same  Hubble type groups. Horizontal lines mark $x_{Y} = 3.4$ (dashed) and 4.3$\%$ (dotted). 
The dispersion in $W_{H\alpha}$ (middle panel) and $x_{Y}$ (right panel) measured at 1 HLR in Sb galaxies is shown as an error bar.
(The corresponding statistical errors are much smaller.) } 
\label{fig:ewha}
\end{figure*}

The above raised issue of the reliability of the recent star formation derived from our optical spectral synthesis analysis is relevant to all our galaxies, not only to early type ones. We now seek for ways to filter out or at least flag objects where \starlight-based SFRs are not reliable enough.

A possible first-cut solution would be to plainly  eliminate all data points where $x_{Y}$ is below, say, twice its uncertainty. Adopting the $\sigma(x_Y) \sim 3\%$ typical uncertainty from the simulations of \citet{cidfernandes14} would then lead to  
a  $x_{Y} > 6\%$  two-sigma criterion to select reliable individual galaxy zones. We note, however, that for statistical reasons a much less restrictive cut would make more sense for the averaging in Hubble type and radial distance bins performed throughout this paper.

We chose to define a criterion based on entirely different precepts. The idea is to use the H$\alpha$ emission equivalent width ($W_{H\alpha}$) to  guide our decision on whether \starlight-derived SFR is indeed tracing recent star formation reliably or not. The rationale goes as follows: 
(1) The recent populations we aim to trace are young enough to photoionize the surrounding gas into HII regions, and hence produce H$\alpha$. (2) Stellar evolution plus straightforward nebular physics  predicts a floor-value of $W_{H\alpha}$ in the range of 1--3 \AA, corresponding to the limit where the interstellar medium is photoionized by hot, old, low mass, evolved stars (HOLMES, as defined by \citealt{FloresFajardo11}. Systems with $W_{H\alpha} \la 3$ \AA\ must therefore have stopped forming stars very long ago, a regime dubbed as ``retired galaxy'' by \citet{stasinska08} and \citet{cidfernandes11}.

This whole scheme is based on the idea that both $x_{Y}$ and $W_{H\alpha}$ trace recent star formation, a corollary of which is that they are correlated. This expectation is fully born out by our data, as seen in the left panel of Fig.\ \ref{fig:ewha}.
The plot shows a (log-scale) density map of $W_{H\alpha}$ versus $x_Y$ for 11894 radial points where H$\alpha$ emission could be measured. White circles trace the mean $W_{H\alpha}$ values in $x_{Y}$ bins, and the grey line shows the corresponding linear fit. We point out that, although expected, this empirical correlation is in no sense tautological, since the two axis are derived from completely independent observables\footnote{Strictly speaking $W_{H\alpha}$ does depend on the \starlight\ run, since the line flux is measured over the residual spectrum obtained after subtracting the \starlight\ fit, but this is only a ``second order'' dependence.}.
In fact, we regard this independence as an added benefit of our approach.

Points with $W_{H\alpha} <  3$ \AA\ in the left  panel of Fig.\ \ref{fig:ewha} have on average $\overline{x}_Y = 3.4 \%$. This  limit on $W_{H\alpha}$ is based both on the observed bimodal distribution of $W_{H\alpha}$ in local Universe galaxies and on long-known theoretical expectations \citep{cidfernandes11}\footnote{
The $W_{H\alpha}$ values expected for galaxies where HOLMES dominate the ionizing flux is in the 0.5--2.4 \AA\ range  (\citealt{binette94}, \citealt{cidfernandes11}, \citealt{gomes15b}). Our $W_{H\alpha} < 3$ \AA\ limit adds a (small) safety cushion to this prediction.
}.
\citet{sanchez15}  propose a more stringent $W_{H\alpha} > 6$ \AA\ cut to isolate regions ionized  by young stars. The mean $x_Y$ for populations with weaker $W_{H\alpha}$ is 4.3$\%$, very close to that obtained with the \citet{cidfernandes11} criterion. Dashed and dotted lines mark $(W_{H\alpha},x_{Y})  = (3 {\rm\, \AA}, 3.4\%)$ and $(6 {\rm\, \AA}, 4.3\%)$ in Fig.\ \ref{fig:ewha}, respectively. It is clear that adopting either of these  cuts should not lead to significantly different results. 

Before elaborating more on the effects of $W_{H\alpha}$ and $x_{Y}$-based reliability criteria, let us first examine how these two properties vary across the face of galaxies.

\subsubsection{The radial profiles or $W_{H\alpha}$ and $x_{Y}$}

The middle and right panels in Fig.\ \ref{fig:ewha} show the average radial profiles of $W_{H\alpha}$ and $x_{Y}$ for the seven morphological bins. Dashed and dotted lines mark the same $(W_{H\alpha},x_{Y})$ limits as in the left panel. 
As in our previous papers (e.g., \citealt{gonzalezdelgado14a}), these average profiles are constructed by first expressing the radial distance for each galaxy in units of the corresponding half light radius (HLR), defined as the length of the elliptical aperture along the major axis that contains half of the total flux at 5635 \AA\ (rest-frame) within the field of view of PPaK. 

The vertical ordering of Hubble types in the middle and right panels of Fig.\ \ref{fig:ewha} follows the expected tendency, with late type systems being more star-forming than early type ones. 
Focusing on the lower part of the plots, we see that E and S0 have mean $W_{H\alpha} < 3$ \AA\ at all locations, confirming that the extended H$\alpha$ emission in these systems is consistent with being produced from photoionization by old stars \citep{sarzi06, kehrig12, papaderos13, singh13, gomes15b}. Whatever little star formation remains in these early type galaxies it is the exception, not the rule. Furthermore, such residual star formation would be located towards the outskirts of these galaxies, as indicated by the rise in their $x_Y(R)$ profiles, reaching 3--5\% for $R  \ga 2$ HLR).

Moving to Sa galaxies, we see that, on average, they have $W_{H\alpha} > 3$ \AA\ at all radii, although they get close to this limit in their central regions (probably reflecting a contribution from retired bulges). Also, except for the central 0.5 HLR, $x_{Y}(R)$ values are all above the  3.4\% line. Beyond 1 HLR their mean $W_{H\alpha}$ oscillates around 6 \AA, so a $< 6$ \AA\ cut would remove significant portions of their disks.
Finally, Fig.\ \ref{fig:ewha} shows that whichever reliability cut we chose to apply would make little difference for Sb and later type galaxies.

In what follows we shall give more emphasis to results obtained by applying a $x_{Y} > 3.4\%$ cut when computing SFR through equation \ref{eq:SFR}, but results obtained with the alternative $W_{H\alpha} > 3$ or 6 \AA\  criteria will also be presented for completeness.
In the next section we explore the impact of these three different criteria on the galaxy-wide SFR and the correlation between $M_\star$ and SFR.

\section{The global main sequence of  star forming galaxies}
\label{sec:GlobalMSSF}

\begin{figure*}
\includegraphics[width=\textwidth]{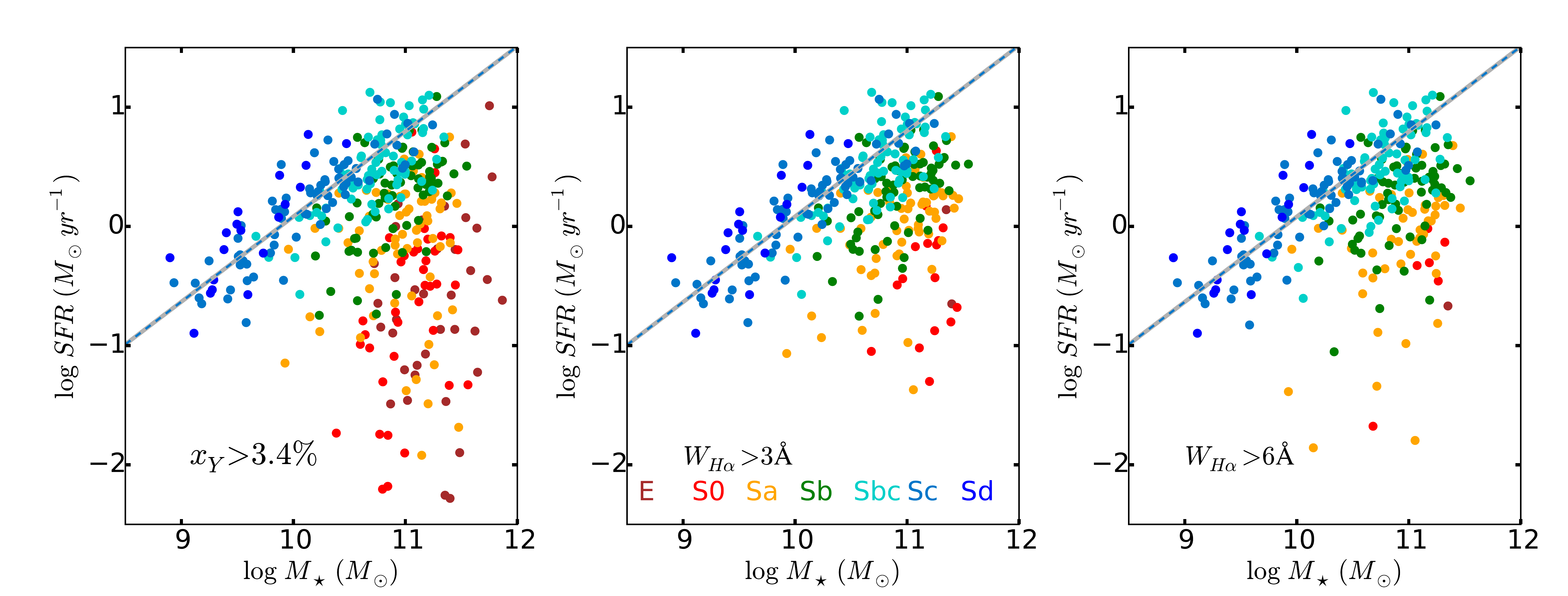}
\caption{Relation between SFR and stellar mass for 416 CALIFA galaxies, color coded by Hubble type. A linear fit to the points of Sc galaxies is shown as grey-blue dashed line. Different panels show the results  obtained considering different selection criteria imposed upon the individual  $xy$ spaxels included in the computation of  the galaxy's total ${\rm SFR} = \sum_{xy} {\rm SFR}_{xy}$: Only $x_{Y} > 3.4\%$ (left), $W_{H\alpha} > 3$ \AA\ (middle), and $W_{H\alpha} > 6$ \AA\  (right).} 
\label{fig:SFMSglobal}
\end{figure*}

As reviewed in the introduction, the main sequence of  star forming galaxies (MSSF) is the name given to the correlation between SFR and $M_\star$ \citep{noeske07}. This correlation has been found in star forming galaxies of the local Universe \citep{brinchmann04}, and seen to persist at least to redshift  $\sim$ 4 \citep{peng10, wuyts11}.
The logarithmic slope of the relation  varies in the range from 0.4 to 1, depending on the galaxy selection criteria and on the indicator used to estimate the SFR \citep{speagle14}. 
Recently, \citet{renzinipeng15} proposed to characterize the main sequence by the ridge line of the star forming peak in a 3D SFR-$M_\star$-Number plot obtained with the SDSS sample. Their objective definition leads to a best fit line given by $\log {\rm SFR} (M_\odot\,{\rm yr}^{-1}) = (0.76 \pm 0.01) \times \log M_\star (M_\odot) - (7.64 \pm 0.02)$. 

Fig.\ \ref{fig:SFMSglobal} shows three versions of the $\log$ SFR vs. $\log M_\star$ relation obtained with our data and methods. 
We call this relation the "global MSSF", in contrast to the "local MSSF" where SFR and $M_\star$ values are replaced by their respective surface densities (cf. Sec.\ \ref{sec:localMSSF} below and Cano-D\'\i az et al.\ 2016). 
The total SFR is calculated for each galaxy using Eq.\ \ref{eq:SFR} and adding the contribution of all spaxels that verify  $x_{Y} > 3.4\%$ (panel a), $W_{H\alpha} > 3$ \AA\ (b), or  $W_{H\alpha} > 6$ \AA\ (c). Galaxies are color coded according to their morphology. 

The dashed gray-blue lines in all panels show $\log {\rm SFR} = a \log M_\star + b$ fits obtained for Sc galaxies.  
The correlation is very similar in the three panels, with a logarithmic slope  $a = 0.77$ and zero point of $b = -7.66$. These values are indistinguishable from those obtained by \citet{renzinipeng15} for the whole SDSS sample. This coincidence is not surprising because Sc, along with Sbc, are the galaxies that contribute the most to the local star formation rate density (Sec.\ \ref{sec:localSFRdensity}), being the ones that produce the ridge line in the MSSF relation. 

As is clear from Fig.\ \ref{fig:SFMSglobal}, the spread in SFR at fixed $M_\star$ is related to galaxy morphology. Table \ref{tab:GlobalMSSF}  lists the slopes and zero points obtained for subsamples of fixed Hubble type. The slopes steepen systematically from 0.34 for Sa to 0.94 for Sd galaxies.
This range is essentially the same as the 0.4--1 quote by \citet{speagle14} as resulting from different selection criteria.
The flattening for the early types also explains why many works obtain a flattening of the MSSF relation at increasing $M_\star$ \citep[e.g.][]{brinchmann04, peng10}. It is clear in Fig.\ \ref{fig:SFMSglobal} that the bending in the main sequence, at least in our sample, is produced by the inclusion of large bulges, such as those in Sa and S0, and also E, where the star formation is already quenched or in the process of being quenched. These galaxies (Sa, S0, and E) are the most massive ones in our sample, but they contribute little to the cosmic star formation (as we will see in Sec.\ \ref{sec:localSFRdensity}), as they are clearly off below the MSSF.

Fig.\ \ref{fig:SFMSglobal} shows that the three alternative cuts defined in the previous section produce practically identical MSSF when galaxies later than Sa are considered. The differences in SFR between the panels become significant in the high $M_\star$ and low SFR regime typical of early type galaxies. Note that the masses are the same from panel to panel, since all spaxels contribute to $M_\star$. What changes is the list of spaxels entering the computation of SFR of each galaxy, hence the differences in the total rate. In practice we obtain a more extended quenched cloud in the left than in middle and right panels. This happens because the  $W_{H\alpha}$-based cuts eliminate most E and several S0 and Sa galaxies altogether, while $x_{Y} > 3.4\%$ is not as restrictive. This again suggests that our estimation of the SFR in E and S0 is uncertain, and our method only provides an upper limit to the real SFR.

\label{sec:globalMS}
\begin{table}
\caption{Parameters of $\log {\rm SFR}
(M_\odot\,{\rm yr} ^{-1}) = a \log M_\star  (M_\odot) + b$ fits of the global MSSF for galaxies of different morphologies. They are obtained for panel (a) in Fig.\ \ref{fig:SFMSglobal}. 
For convenience, the corresponding SFR for a $10^{10} M_\odot$ galaxy is also listed (in $M_\odot\,$yr$^{-1}$).
Note that the slope is significantly smaller in Sa that in later spirals because most of the Sa galaxies are off the MSSF.}
\begin{tabular}{lccccc}
\hline\hline
 morph. & Sa & Sb & Sbc & Sc & Sd    \\      
\hline
logarithmic slope     ($a$)      & 0.34    &   0.65  & 0.71  & 0.77    & 0.94     \\
zero-point ($b$)                    & $-3.87$  &  $-6.83$  & $-7.05$  & $-7.66$   & -9.12   \\
$\log {\rm SFR}(M_\star= 10^{10} M_\odot)$ & $-0.47$  &  $-0.33$  & $-0.05$  & $-0.04$   & $0.28$    \\
 \hline\hline
\label{tab:GlobalMSSF}
\end{tabular}
\end{table}

\section{Radial structure of the recent star formation }
\label{sec:RadialStructures}

We now present a series of results related to spatially resolved SFR measurements of CALIFA galaxies. We focus on the radial structure of the star formation rate surface density, $\Sigma_{\rm SFR}$ (also referred to as the intensity of star formation), and the local specific star formation rate, $\Sigma_{\rm SFR}/\mu_\star$. 

Using \pycasso\ we obtain, for each galaxy, 2D maps of the recent SFR computed as in equation \ref{eq:SFR} with $t_{SF} = 32$ Myr.
Each 2D map is then azimuthally averaged to obtain the radial variation of the $\Sigma_{\rm SFR}$. Only spaxels that meet the  criterion of $x_{Y} > 3.4\%$ are included in the azimuthal average. Elliptical apertures $\Delta R = 0.1$ HLR in width are used to extract the radial profiles, with ellipticity and position angle obtained from the moments of the 5635 \AA\ flux image. We express the radial distance in units of HLR to allow comparison of the profiles of individual galaxies, and to produce stacks as a function of Hubble type and/or stellar mass.

\subsection{Radial profiles of $\Sigma_{\rm SFR}$ and the role of morphology}

\begin{figure*}
\includegraphics[width=\textwidth]{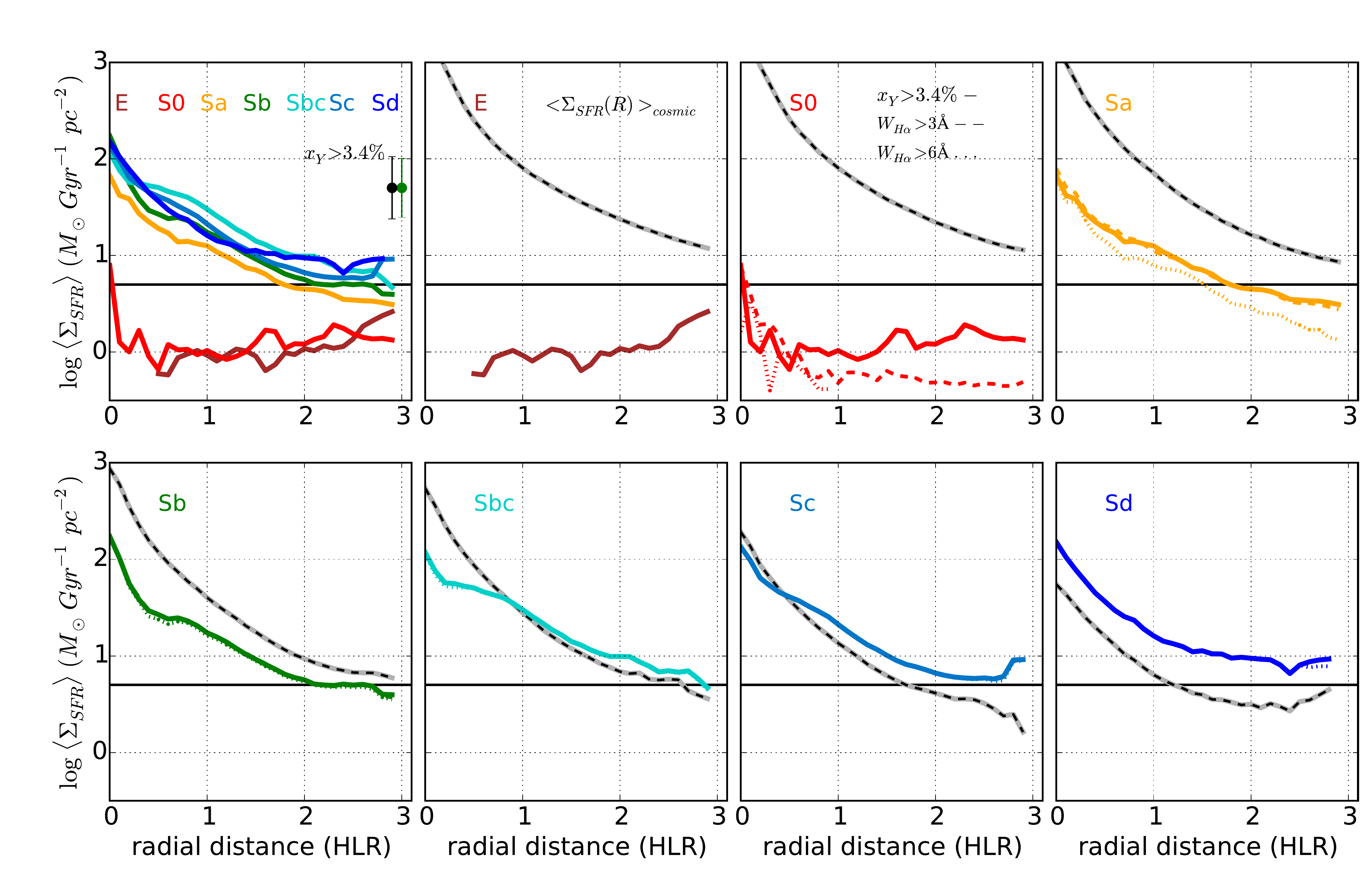}
\caption{ 
{\em Upper left: }  Radial profiles (in units of HLR) of the star formation rate surface density ($\Sigma_{\rm SFR}$), averaged in seven morphology bins. The horizontal line at 5  $M_\odot\,$Gyr$^{-1}\,$pc$^{-2}$ marks the average $\Sigma_{\rm SFR}$ in the Milky Way (MW). Only the locations where the light fraction due to young populations  is higher than 3.4\% are included. Error bars show the dispersion (not the uncertainty) in $\log \Sigma_{\rm SFR}$ for all the spirals (black) and for Sb galaxies (green).  {\em Other panels:} 
For each Hubble type, solid, dashed, and dotted lines show the mean profiles obtained by excluding $x_{Y} < 3.4\%$, $W_{H\alpha} <  3  \AA$, and $W_{H\alpha} <  6  \AA$, respectively. 
The grey-dashed lines show $\langle \Sigma_{\rm SFR}(R) \rangle_{cosmic}$ profiles, the SFR surface density profiles obtained assuming a constant rate of star formation throughout the Hubble time.
} 
\label{fig:SFRSD}
\end{figure*}

Fig.\ \ref{fig:SFRSD} shows azimuthally averaged radial profiles of $\Sigma_{\rm SFR}$ stacked by Hubble type. The upper-left panel shows the results for all the seven morphological classes together.

All spirals show $\Sigma_{\rm SFR}(R)$  decreasing with radial distance, with a typical gradient 
(measured in the central 1 HLR)  $\Delta\log\Sigma_{\rm SFR} = -0.78$ dex/HLR.
Interestingly, the  $\Sigma_{\rm SFR}(R)$ at any radius falls within a relatively tight range of values. At  $R = 1$ HLR our average $\Sigma_{\rm SFR}$ is 20 $M_\odot\,$Gyr$^{-1}\,$pc$^{-2}$, with a dispersion of 0.13 dex between spirals of different Hubble type. 
This is about one to two orders of magnitude smaller than the global $\Sigma_{\rm SFR}$ measured in starbursts and local Lyman break analogs \citep{heckman05}, but consistent with the value obtained by \citet{schiminovich07} for a complete sample of GALEX star forming galaxies\footnote{They divide half of the total SFR (derived from the UV luminosity) by an area equal to $\pi$ HLR$^2$.}.

The plot also illustrates how E and S0 are clearly distinct from the spirals. Their radial profiles are  flat (except for some slight increase at the center of S0). The $\Sigma_{\rm SFR}$ at 1 HLR is $\sim 1 M_\odot\,$Gyr$^{-1}\,$pc$^{-2}$, a 20-fold decrease from spirals (maybe more given that our estimates of SFR for early types are probably upper limits).

Each of the other panels in Fig.\ \ref{fig:SFRSD} shows the radial profile of  $\Sigma_{\rm SFR}$ for each Hubble type, now computed with each of our three reliability cuts. As already discussed, imposing $x_{Y} > 3.4\%$ (solid line),  $W_{H\alpha} > 3$ (dashed) or 6 \AA\ (dotted) makes no difference for galaxies later than Sb, so much so that the three $\Sigma_{\rm SFR}$ profiles are hardly distinguishable. The effects of the somewhat more restrictive $W_{H\alpha}$-cuts start to be noticed in Sa galaxies, become evident  in S0 (factor of $\sim 2$ difference) and grow even larger in E (where the dotted and dashed lines fall off below the plot limits). 
As previously discussed, though some of our E galaxies do exhibit signs of recent star formation \citep{gomes15c}, most can be regarded as quenched systems, which retired from forming stars long ago. We note in passing that although our estimates for E and S0 are very uncertain, the typical value of $\sim 1 M_\odot\,$Gyr$^{-1}\,$pc$^{-2}$ is consistent with the global $\Sigma_{\rm SFR}$ in early type galaxies estimated by \citet{schiminovich07}.

The individual panels of Fig.\ \ref{fig:SFRSD} present two other lines which allow for interesting comparisons. The first is the horizontal line
at $5 M_\odot\,$Gyr$^{-1}\,$pc$^{-2}$, which marks the $\Sigma_{\rm SFR}$ of the Milky Way. This value is obtained by dividing the recent SFR of the Milky Way (MW), $1.6 M_\odot\,$yr$^{-1}$, by the area of a disk of radius $1.2 \times$ the Galactocentric radius of the Sun, 8.33 kpc \citep{licquia15}. This distance is equivalent to $\sim 2$ HLR of typical Sb-Sbc CALIFA galaxies. 
The results in  Fig.\ \ref{fig:SFRSD} suggest that, in the inner $\sim 2$ HLR, most  spirals in the main sequence have $\Sigma_{\rm SFR}$ higher than the average $\Sigma_{\rm SFR}$ in the Milky Way. This is not unexpected because it is known that the SFR in the MW is significantly lower than in other spirals of similar type and mass (see \citealt{kennicuttevans12}, particularly the discussion around their figure 7, where the $\Sigma_{\rm SFR}$ profiles of the Galaxy and NGC 6946 are compared).

The dashed-grey-black lines in Fig.\ \ref{fig:SFRSD} present another useful reference to compare our results to. They represent the 
$\Sigma_{\rm SFR}(R)$ the galaxy should have if it formed stars at a constant rate throughout the lifetime of the Universe. Except for the correction for returned mass (a simple scaling factor under most circumstances), these $\langle \Sigma_{\rm SFR}(R) \rangle_{cosmic}$ profiles reflect the stellar mass surface density profile divided by 14 Gyr.

E and S0 are clearly quenched, with $\Sigma_{\rm SFR} \sim 2$ dex below their $\langle \Sigma_{\rm SFR}(R) \rangle_{cosmic}$. Sa and Sb galaxies, although still active in forming stars,  do so at a lower rate than in the past. In contrast, spirals of later types are forming stars at a rate similar to or  higher than their $\langle \Sigma_{\rm SFR}(R) \rangle_{cosmic}$.  In the central 0.5 HLR of Sbc, presumably their bulges, $\Sigma_{\rm SFR}$ have decreased with respect to the past, although the disk (outside 1 HLR) is currently forming stars somewhat more actively than in the past.  The intersection of the $\Sigma_{\rm SFR}$ and $\langle \Sigma_{\rm SFR}(R) \rangle_{cosmic}$ curves occurs even closer to the nucleus in Sc, while in Sd the current intensity of star formation exceeds the past average at all radii.

\subsection{The dependence of $\Sigma_{\rm SFR}(R)$ on stellar mass}

\begin{figure*}
\includegraphics[width=\textwidth]{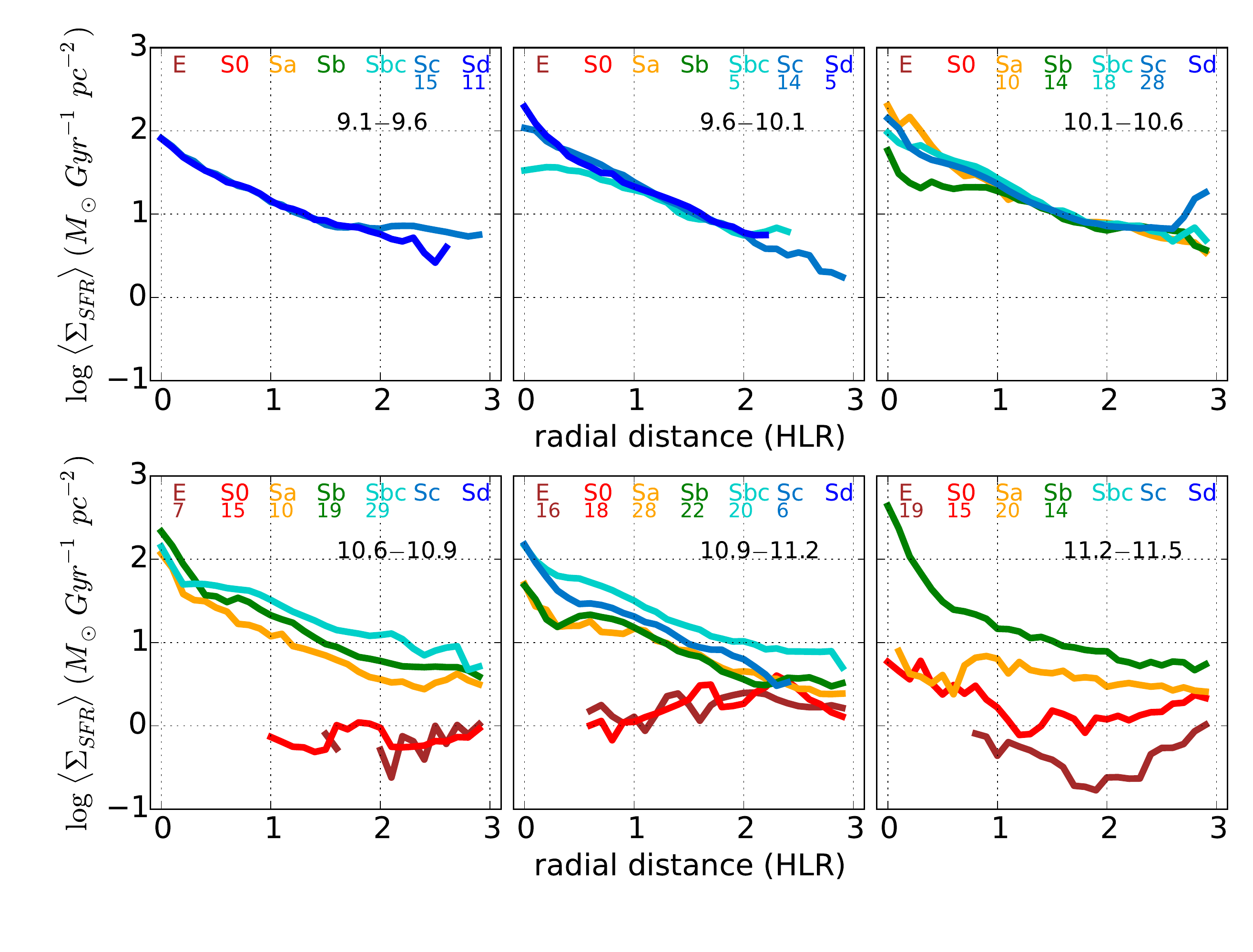}
\caption{Radial profiles of $\Sigma_{\rm SFR}$ for different Hubble types in six galaxy stellar mass bins. From top-left to bottom-right: $\log M_\star (M_\odot) = 9.1$--9.6, 9.6--10.1, 10.1--10.6, 10.6--10.9, 10.9--11.2, 11.2--11.5. In each panel, the average profile for each Hubble type is plotted if more than five galaxies have masses in the corresponding  $M_\star$-bin. The number of galaxies in each bin is also labeled and color coded.
} 
\label{fig:SFRSD_tipo_masa}
\end{figure*}

As usual, it is difficult to disentangle the relative roles of morphology and $M_\star$, but CALIFA has grown large enough a sample to attempt to tackle this issue by plain brute force statistics.

Fig.\ \ref{fig:SFRSD_tipo_masa} shows $\Sigma_{\rm SFR}(R)$  profiles as a function of both $M_\star$ and morphology. Besides the seven Hubble types we now break up the sample in six mass bins: $\log M_\star (M_\odot) = 11.5$--11.2, 11.2--10.9, 10.9--10.6, 10.6--10.1, 10.1--9.6, and 9.6--9.1. In each panel (one per $M_\star$ bin), the average profile for each Hubble type is plotted if it contains more than five galaxies. These plots allow to evaluate how $\Sigma_{\rm SFR}(R)$ changes with Hubble type for galaxies of similar mass.

An inspection of Fig.\ \ref{fig:SFRSD_tipo_masa} shows that spirals with  $M_\star \la 4 \times 10^{10} M_\odot$   (top panels) have very similar $\Sigma_{\rm SFR}$ profiles. When relevant, the differences occur in the inner regions. Above this mass, the profiles start to disperse, although they are still packed in a relatively small range of $\Sigma_{\rm SFR}$ values. 
This high degree of uniformity  is a remarkable result taking into consideration that the sample covers all types of spirals and two orders of magnitude in galaxy mass. In Section \ref{sec:Local2Global} we speculate that this behaviour is intimately linked to the tightness of the MSSF.

In contrast, E and S0 have 
$\Sigma_{\rm SFR}$ profiles well below those in spirals of similar mass. This suggests that in massive galaxies with a large spheroidal component the star formation is significantly quenched in the whole galaxy.
 However, this effect seems to be more relevant in the centers than in the outskirts, as suggested by the flat profiles in E and S0 in comparison with the radially decreasing $\Sigma_{\rm SFR}(R)$ profile in spirals. 
Note that the most massive Sa galaxies in the sample show a bimodal behavior, with a smooth decrease of  $\Sigma_{\rm SFR}(R)$ outwards of 1 HLR, and a relative rate in the central part that is almost flat and significantly depressed with respect to spirals of later types. Again, this points out  to the relevance that the formation of a big bulge may have in quenching the star formation in galaxies.

\subsection{Radial structure of the local specific star formation rate}

\begin{figure}
\includegraphics[width=0.5\textwidth]{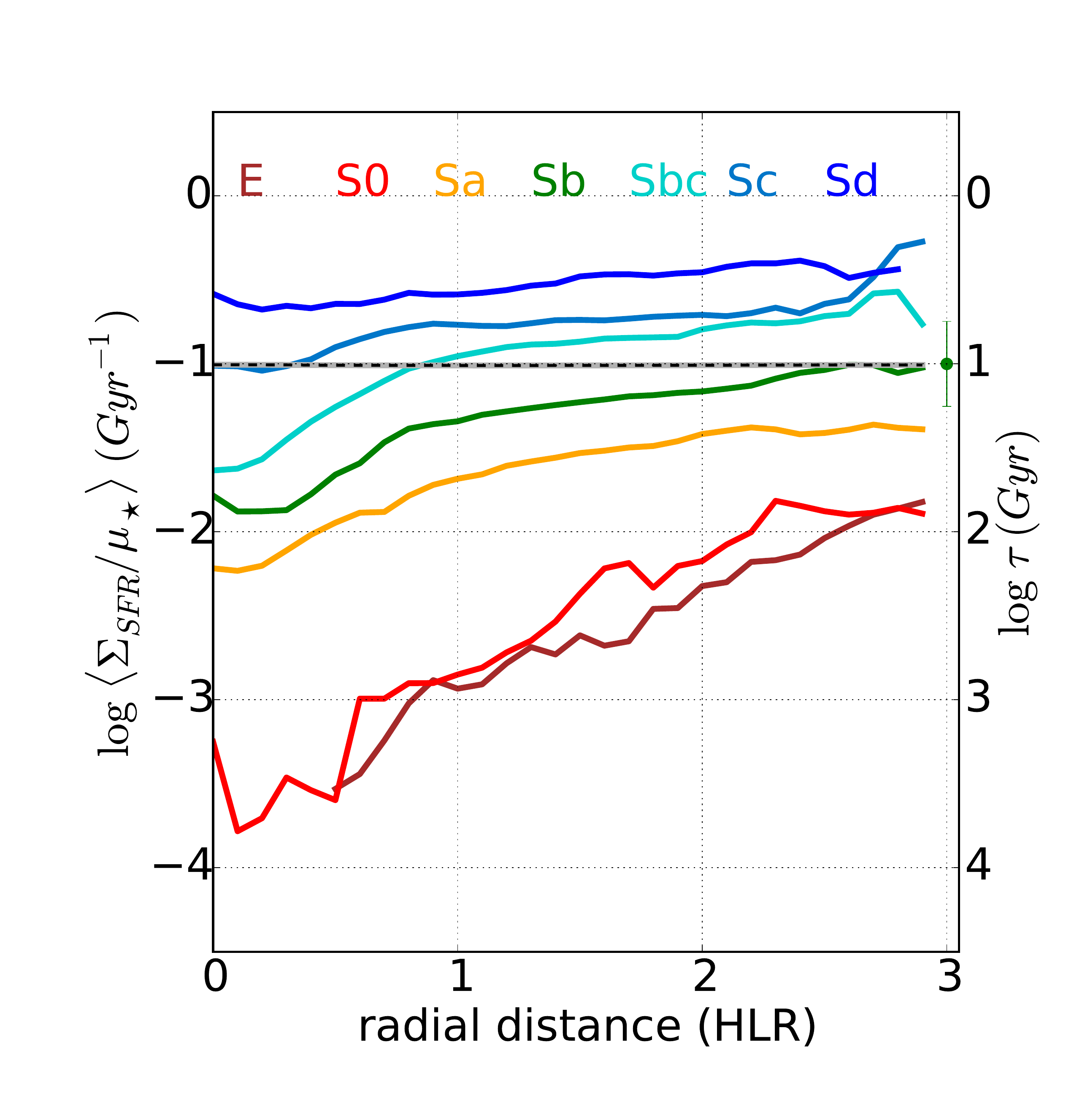}
\caption{As the upper-left panel of Fig.\ \ref{fig:SFRSD}, but for the local specific star formation rate, ${\rm sSFR}(R) = \Sigma_{\rm SFR}(R) / \mu_\star(R)$. The error bar shows the  $1\sigma$ dispersion in $\log {\rm sSFR}$ for Sb galaxies.
The grey-dashed line at  ${\rm sSFR} =0.1\,$Gyr$^{-1}$ marks the value adopted by \citet{peng10} as a threshold to separate star forming galaxies from quiescent systems.  }
\label{fig:sSFR}
\end{figure}

For a galaxy, the specific star formation rate is defined by ${\rm sSFR} = {\rm SFR}/M_\star$. Overlooking trivial multiplicative factors (see equation \ref{eq:b}), it gives a measure of the relative rate at which stars are forming now in a galaxy with respect to the past average rate. Because the relation between SFR and $M_\star$ is sub-linear (e.g. Fig. \ \ref{fig:SFMSglobal}), the sSFR declines with  galaxy mass. Also, because of the tightness of the MSSF relation, star forming galaxies occupy a correspondingly tight locus in the sSFR vs.\ $M_\star$ space, but bulge dominated galaxies display a much larger spread of sSFR at a fixed galaxy mass (\citealt{schiminovich07, salim07}---see also Fig. \ \ref{fig:SFMSglobal})

In analogy with the global sSFR, CALIFA data allow us the study of the {\em local} sSFR, defined by the ratio $\Sigma_{\rm SFR}/\mu_\star$, that measures the relative rate of ongoing star formation with respect to the past in each position in a galaxy.

Fig.\ \ref{fig:sSFR}  shows the results of stacking the ${\rm sSFR}(R) = \Sigma_{\rm SFR}(R) / \mu_\star(R)$ profiles by Hubble type. 
These  profiles show a clear ranking with  morphology, increasing from early to late Hubble type.  We obtain, at $R = 1$ HLR, $\log {\rm sSFR} ({\rm Gyr}^{-1}) = -2.94$, $-2.85$, $-1.68$, $-1.34$, $-0.95$, $-0.77$, and $-0.59$ for  E, S0, Sa, Sb, Sbc, Sc, and Sd bins, respectively.
This ordering is preserved at any given radial distance, as is also the case with other stellar population properties such as mean stellar age, metallicity, and $\mu_\star$ \citep{gonzalezdelgado15}.

Fig.\ \ref{fig:SFRSD} showed that  $\Sigma_{\rm SFR}(R)$ profiles are very similar for all spirals, so the scaling of sSFR$(R)$ seen in Fig.\ \ref{fig:sSFR} is a direct consequence of the variation of $\mu_\star(R)$ with Hubble type, increasing from $\mu_\star(R = 1\,{\rm HLR}) \sim 100$ to $1000\, M_\odot\,$pc$^{-2}$, from Sd to Sa galaxies.
The opposite happens for early type galaxies, with the sSFR$(R)$ profiles of E and S0 galaxies running well below those of Sa, while their $\mu_\star(R)$ profiles are similar \citep{gonzalezdelgado15}. The difference in this case comes from the much smaller levels of star formation in these systems.
  
All the galaxies have outwardly increasing sSFR profiles. Fig.\ \ref{fig:sSFR} shows that in spirals sSFR$(R)$ grows faster {with radius  in the inner 1 HLR than outwards, probably signaling the bulge-disk transition. Assuming that the central 0.1 HLR is  dominated by the bulge and that the disk dominates outside 1 HLR, we can compare the sSFR values in these two morphological components through a ratio like
${\rm sSFR}(R = 0.1) / {\rm sSFR}(R = 1.5)$.
For early type spirals (S0, Sa, Sb) this exercise results in that the sSFR of bulges is on average 0.40 dex smaller than in the disks.
The difference is larger, 0.60 dex, for Sbc, while for later later types (Sc, Sd, with their small on non-existent bulges) it decreases to 0.24 dex.
The sample dispersions around these values is $\sim 0.3$ dex.

As in the case of the global sSFR, the local one  can also be expressed as a characteristic time-scale of star formation,  $\tau(R) =  {\rm sSFR}(R)^{-1}$ that,  independently of IMF and cosmology,  tells the period of time that the system  needs to build its current stellar mass forming stars at the present rate\footnote{
This standard reading of $\tau = {\rm sSFR}^{-1}$ actually neglects the difference between the mass turned into stars and that which stays in stars (or remnants). Because the denominator in sSFR is the current stellar mass, a rigorous definition would require a $(1 - {\cal R})^{-1}$ correction for the returned mass fraction ${\cal R}$, not important for the discussion at this point. 
}.
Measured at $R = 1$ HLR, $\tau$ ranges from 12.6 Gyr in Sbc to 5 Gyr in Sd galaxies.
Early type spirals (Sa, Sb), S0, and E would all need more than the Hubble time to build their mass at their current SFR.

The dashed grey-black line in Fig.\ \ref{fig:sSFR}  at  ${\rm sSFR} =0.1\,$Gyr$^{-1}$ marks the value adopted by \citet{peng10} as a threshold to separate star forming galaxies from quiescent systems. This line also marks the sSFR that galaxies should have to build their mass at the present rate during a Hubble time (approximated to 10 Gyr). The comparison of  0.1 Gyr$^{-1}$ with the sSFR profiles indicates 
that Sd, Sc, and the disks of Sbc are very actively forming new stars, while Sa and Sb galaxies and  the bulges of Sbc, although still forming stars, are evolving to quiescent systems.

Finally, E and S0 have sSFR$(R)$ values 10--100 times smaller than 0.1 Gyr$^{-1}$, with a steep increase outwards. This suggests that quenching in these galaxies has progressed inside-out.

\section{Discussion}
\label{sec:Discussion}

The central goal of this paper, the one embodied in its very title, was fulfilled in the previous section with the results on the radial profiles of $\Sigma_{SFR}$ and sSFR for galaxies along the Hubble sequence (Figs.\ \ref{fig:SFRSD} to \ref{fig:sSFR}). In this final part we go beyond this point and examine a few related issues. First we take advantage of the volume corrections computed by \citet{walcher14} to extrapolate the SFR computed from our sample to a local Universe SFR density and how it breaks up into contributions from different Hubble types and radial regions (Section \ref{sec:localSFRdensity}).  Secondly we re-express our results  (both the radial profiles and the local Universe average) for SFR in terms of the birthrate parameter $b$ (Section \ref{sec:bScalo}). We then turn our eyes to the spatially resolved version of the global MSSF. Like the SFR and $M_\star$ for entire galaxies (Fig.\ \ref{fig:SFMSglobal}), their surface densities $\Sigma_{SFR}$ and $\mu_\star$ correlate strongly, with a morphology-related scatter (Section \ref{sec:localMSSF}). Finally, we gather our results to formulate an empirical scenario which identifies the origin of the global MSSF (Section \ref{sec:Local2Global}).

\subsection{The SFR volume density in the local Universe}
\label{sec:localSFRdensity}

\begin{figure}
\includegraphics[width=0.5\textwidth]{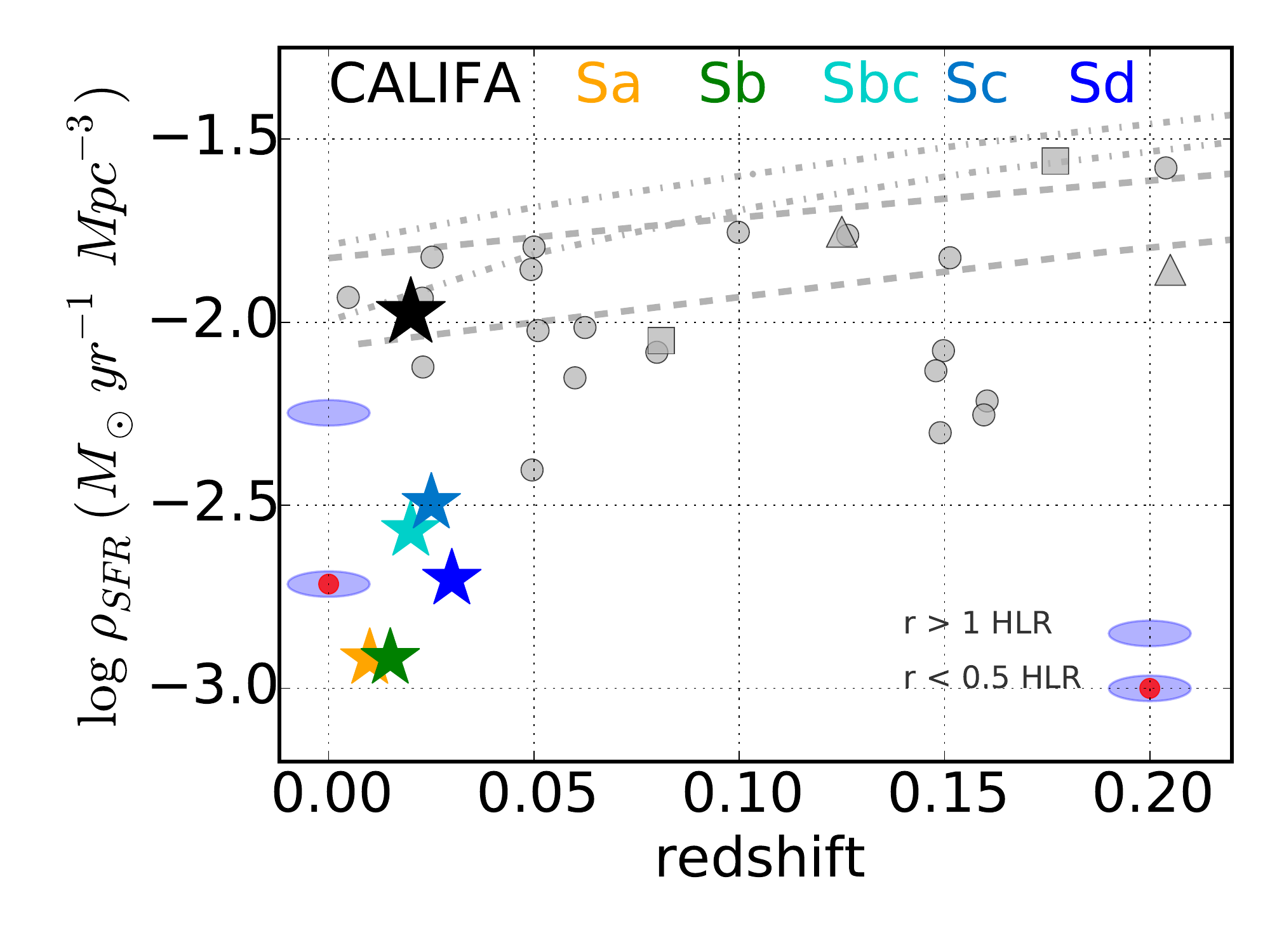}
\caption{The star formation rate density in the present study (black star). Colored stars represent the contribution of each morphological type to the total star formation rate density ($\rho_{\rm SFR}$). The blue ellipse represents the contribution to $\rho_{\rm SFR}$ of the regions outside 1 HLR; and the red-dot blue ellipse represents the contribution to $\rho_{\rm SFR}$ of the regions inside 0.5 HLR. Other results are from recent determinations by \citet{gunawardhana13, gunawardhana15} (grey triangles) and their compilation (grey points), and the redshift evolution of $\rho_{\rm SFR}$ from \citet{hopkinsbeacom06} (two top grey-dotted lines are $\pm 1 \sigma$ of their relation);  \citet{madau14} (middle grey-dashed line); \citet{fardal07} (bottom grey-dashed line); and from the fossil record method applied to SDSS data by \citet{panter03} (grey squares). When needed, literature values have been scaled to a Salpeter IMF.
} 
\label{fig:sfrdensity}
\end{figure}

CALIFA, as many other samples, is not volume-limited, but  can be ``volume-corrected" using the $V_{max}$ method \citep{schmidt68}. $V_{max}$ is the  volume available per galaxy,  here calculated for a diameter-limited sample by assuming that the ratio between apparent and linear isophotal size of a galaxy depends only on its angular diameter distance (see \citealt{walcher14} for details). 

We use this method to extend our results to those expected for local Universe galaxies as a whole. In particular, we transform our SFR estimates into the volume density of SFR, $\rho_{\rm SFR}$, by  adding ${\rm SFR} / V_{max}$ for our galaxies and correcting the result by 
$\times 937/414$, the ratio of galaxies in the mother sample to those used in this paper\footnote{For this analysis we exclude NGC 4676B and NGC 5947 because they do not belong to the original CALIFA sample, and have no associated $V_{max}$ estimates.}.
CALIFA is a local sample, so there is no need to correct for evolution over the lookback time spanned by its redshift limits. 

This process yields  $\rho_{\rm SFR}= 0.0105 \pm 0.0008$ (random) $M_\odot\,$yr$^{-1}\,$Mpc$^{-3}$.  Fig.\ \ref{fig:sfrdensity} places our estimate (black star) in the $\rho_{\rm SFR}$ vs.\ $z$ diagram, along with other values from the literature, coming from different samples and methods. The dashed lines shows the evolution of $\rho_{\rm SFR}$ from \citet{madau14}, \citet{hopkinsbeacom06}, and \citet{fardal07}. We also include the local $\rho_{\rm SFR}$ from the compilation of \citet{gunawardhana15,gunawardhana13}, and the results obtained by \citet{panter03} from the fossil record method applied to the SDSS data. 
When necessary, the literature results are scaled to a Salpeter IMF.  Our estimate is smaller by 0.15 dex and  higher by 0.05 dex than the values at $z=0$ from \citet{madau14} and \citet{fardal07}, respectively. 
It is also in excellent agreement with the $z< 0.1$ estimates compiled by \citet{gunawardhana15,gunawardhana13}, which average to
$0.0109 M_\odot\,$yr$^{-1}\,$Mpc$^{-3}$.

Obviously, the above refers to integrated measurements, which in our case tantamounts to collapsing all our 11894 radial points  into a single number. To better explore our data  Fig.\ \ref{fig:sfrdensity} also shows the contribution to the overall  $\rho_{\rm SFR}$ from the different morphological types, plotted as stars (color coded by their morphology). 
It is clear that Sbc, Sc, and Sd galaxies dominate the $\rho_{\rm SFR}$ budget. 
Together they contribute  $\sim 75\%$ of $\rho_{\rm SFR}$,  despite accounting for only $\sim 24\%$ of the stellar mass volume density of the local Universe ($\rho_\star$, computed following the same methodology).
In contrast, Sa and Sb galaxies contribute  $\sim 22\%$ to  $\rho_{\rm SFR}$  and $\sim 33\%$ to $\rho_\star$, while  E and S0 add less than  2\% to  $\rho_{\rm SFR}$  but 43\% to $\rho_\star$.

In terms of spatial origin, 53\% of $\rho_{\rm SFR}$ comes from the regions outwards of 1 HLR, 29\% from $0.5 < R < 1$ HLR, and 18\% from the inner 0.5 HLR. In contrast, the $\rho_\star$ budget for these same regions are 40, 25 and 35\%, respectively. 
Most of the ongoing star formation thus occurs outside the centers, in disk  dominated regions, while the stellar mass is more evenly distributed with radius. If we take as reference the half mass radius (HMR), which is typically $0.8 \times$ the HLR \citep{gonzalezdelgado15}, we find that only 35$\%$ of $\rho_{SFR}$ comes from the regions inside the central 1 HMR, suggesting again that most of the star formation density comes from the disk dominated regions.

\subsection{The birthrate parameter}
\label{sec:bScalo}

\begin{figure}
\includegraphics[width=0.5\textwidth]{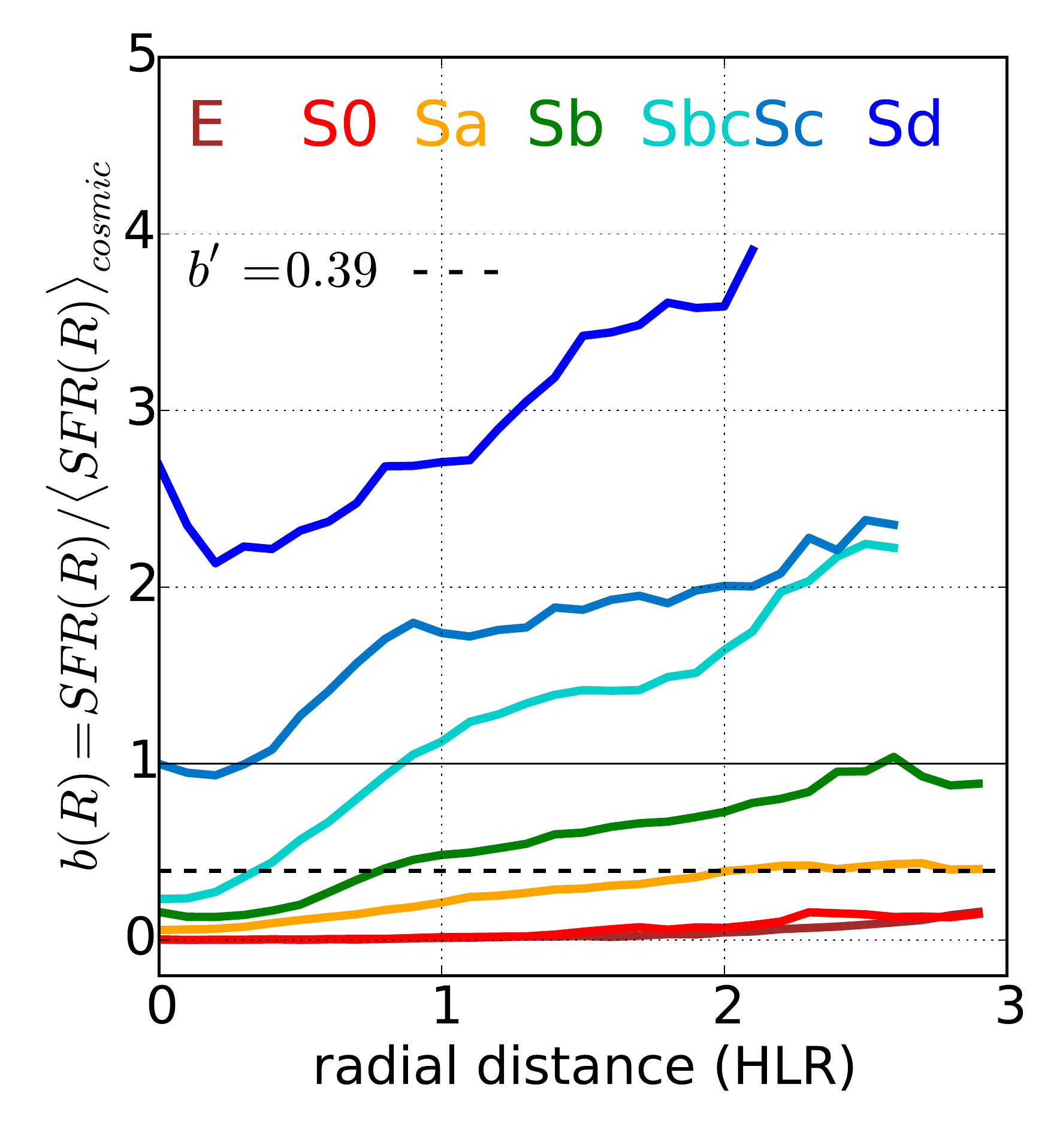}
\caption{Radial profiles of the birthrate parameter, $b(R)$,  averaged in seven Hubble type bins;
$b(R)$ compares the present to the past-average SFR at radius $R$ (cf.\ equation \ref{eq:bradial}).
The dashed and solid black lines mark, respectively, our (volume-corrected) value for the local Universe, and the reference value $b(R) = 1$.
} 
\label{fig:bparameter}
\end{figure}

It is often useful to consider the SFR in relation to some fiducial value, instead of in absolute units. A classical example is the birthrate parameter, $b$, that measures the current SFR of a system with respect to its lifetime average, $\langle {\rm SFR} \rangle_{cosmic}$ \citep{kennicut83, scalo86}. 
This parameter conveniently separates galaxies with declining SFRs ($b < 1$) from those with SFR increasing ($b > 1$) from past to present.
$b$ and  sSFR are related by

\begin{equation} 
\label{eq:b}
b =   {\rm SFR} \, / \, \langle {\rm SFR} \rangle_{cosmic}  =  {\rm sSFR} 
   \ \ t_\infty  \ (1-\cal R) 
 \end{equation}

\noindent where ${t_\infty}$ is the time over which the galaxy has formed stars\footnote{${t_\infty} = t_{H}(z) - t_{form}$, where $t_{H}(z)$ is the Hubble time at redshift $z$ and $t_{form}$ is the time of formation.}, and $\cal R$ denotes the fraction of the 
mass initially turned into stars which is returned to the interstellar medium by stellar evolution.
In practice, $b = 10.08\,  {\rm sSFR}({\rm Gyr}^{-1})$ for $t_\infty = 14$ Gyr and (1-$\cal R$) = 0.72 (average over our sample and the Salpeter IMF assumed in the models).

 A volume-corrected value of $b$ representative of the local Universe can be obtained from

\begin{equation} 
\label{eq:bradial}
b^\prime = \frac{ \sum_i  {\rm SFR}_i  V_{max,i}^{-1} }{ \sum_i  \langle {\rm SFR} \rangle_{cosmic, i} V_{max,i}^{-1} }
\end{equation}

We find that $b^\prime = 0.39 \pm 0.03$ (random). Put in words, the present-day Universe is forming stars at a little over 1/3 of its past average rate.

As for $\rho_{\rm SFR}$ (Section \ref{sec:localSFRdensity}), though useful, this one-number-summary of the star formation history of the Universe as a whole averages over the richness of information in CALIFA data. Our spatially resolved observations allows for definitions of $b$ that take into account its variation within galaxies \citep{cidfernandes13}, the simplest of which is  

\begin{equation} 
\label{eq:bradial}
b(R) = \Sigma_{\rm SFR}(R) \, / \, \langle  \Sigma_{\rm SFR}(R) \rangle_{cosmic} 
\end{equation}

\noindent This radial profile $b(R$) behaves exactly as the  global $b$ of equation \ref{eq:b}. There is, however, a relevant assumption implicit in this comparison of past and present as a function to radius, namely, that (statistically) stars do not move too far from their birthplaces along their lives.

Fig.\ \ref{fig:bparameter} shows our results for $b(R)$ for our seven Hubble types.
Clear and systematic trends are identified with both radial distance and morphology. First, $b(R)$ increases outwards, as expected from the inside-out growth of galaxies \citep{perez13, gonzalezdelgado14a, sanchez-blazquez14, sanchez14, gonzalezdelgado15}. Secondly, $b(R)$ scales in amplitude with Hubble type, increasing from early to late spirals. Sd and Sc galaxies are currently forming stars faster than in the past at all radii. The disks ($R > 1$ HLR) of Sbc galaxies also show $b(R) > 1$, but their bulges are forming stars at lower rates than in the past. 
Sb and earlier types have $b(R) < 1$ throughout their disks and bulges.

Finally, we note that spheroids (E, S0, and  the inner regions of early type spirals, presumably associated to bulges) all have $b < b^\prime$. Star formation has thus stopped (or been quenched) sometime ago. Most regions in Sa have $b(R) < 0.39$, so, even though these galaxies are still forming stars, they are located in the transition between the MSSF and the quenched cloud.

\subsection{The local  main sequence of star formation}
\label{sec:localMSSF}

\begin{figure*}
\includegraphics[width=0.5\textwidth]{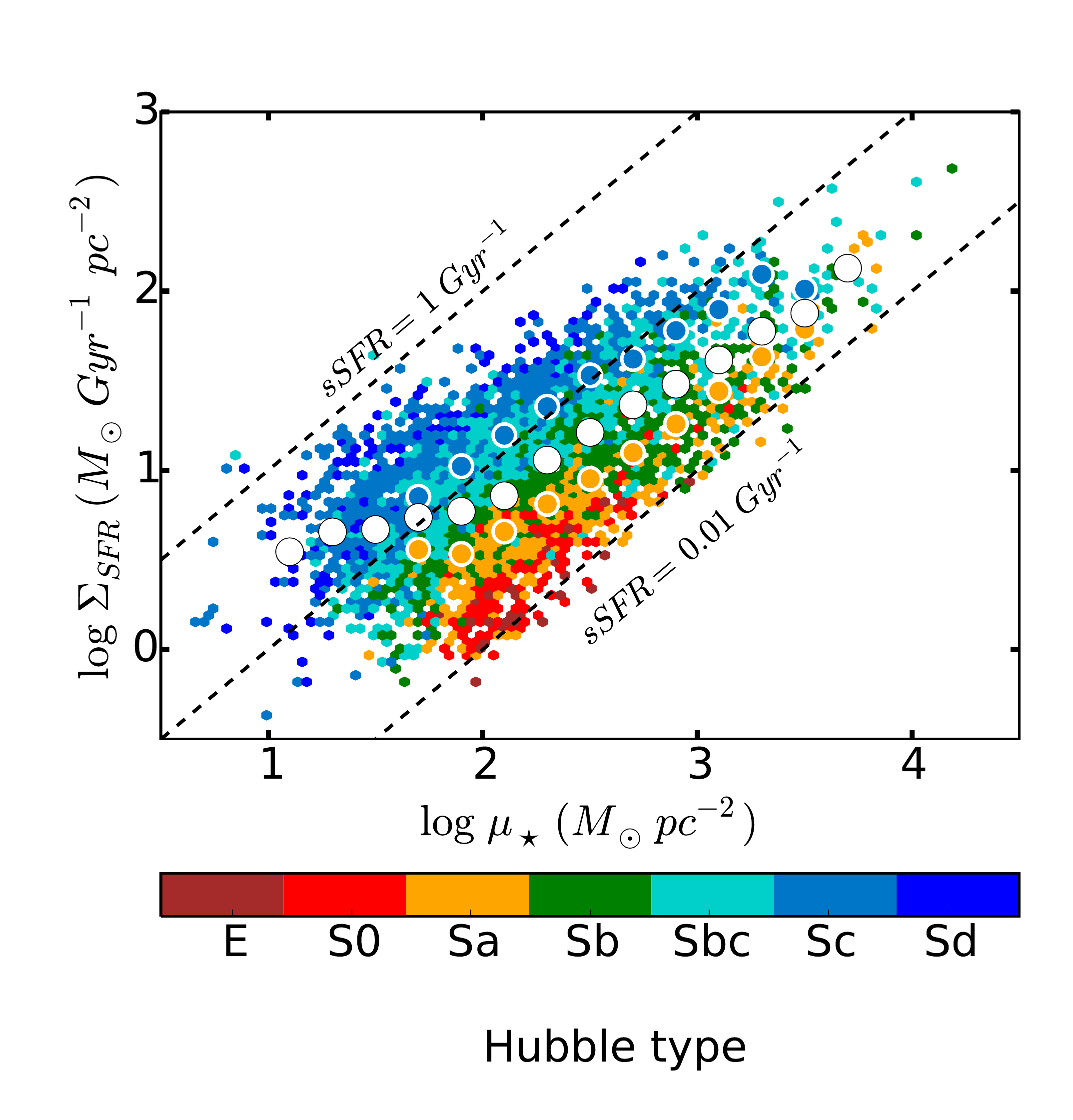}
\includegraphics[width=0.5\textwidth]{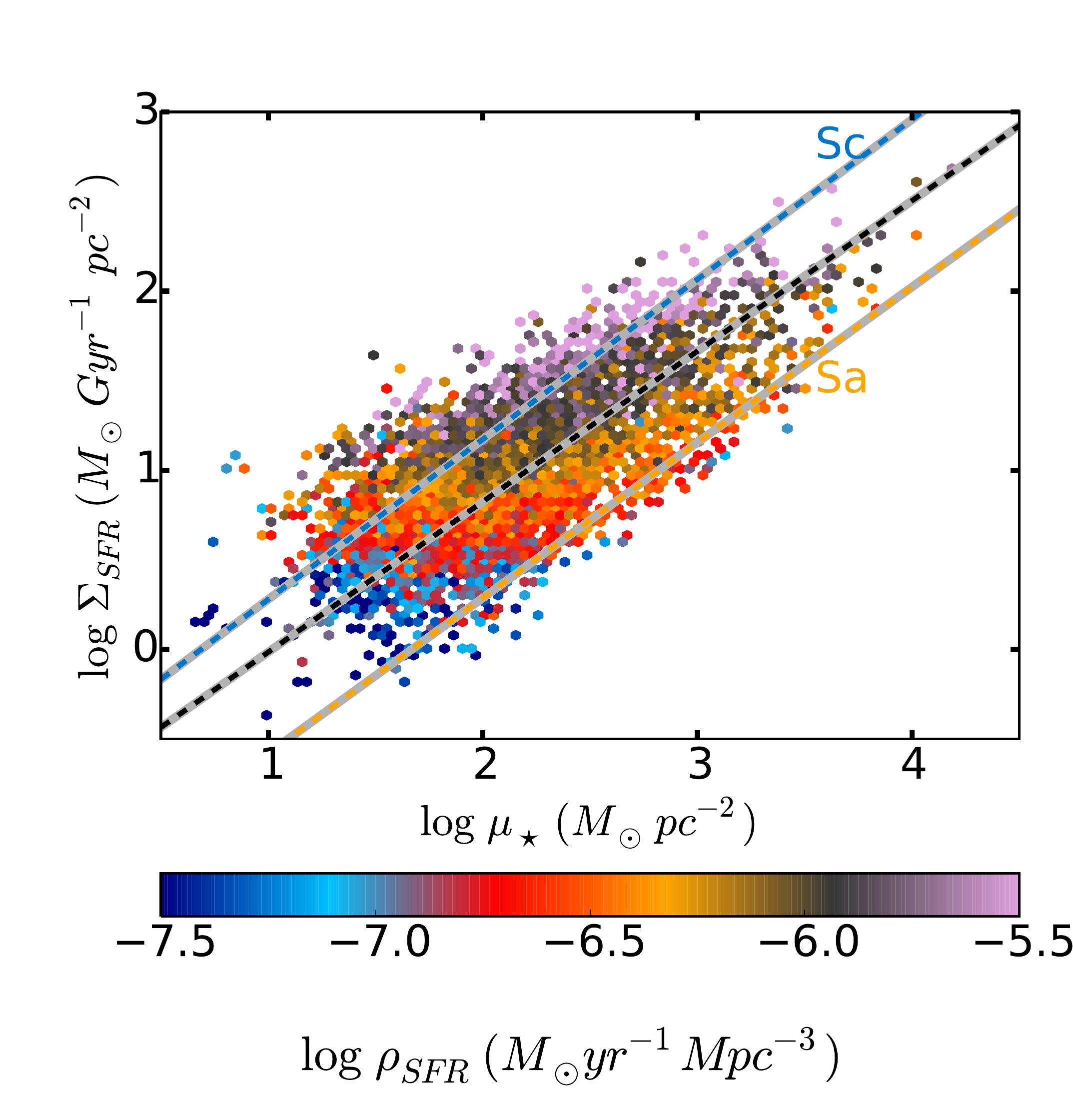}
\caption{
{\em Left:} The local star forming main sequence $\mu_\star$-$\Sigma_{\rm SFR}$ relation, color coded by morphology. Large white circles show the mean (i.e., $\mu_\star$-binned) relation for all points; orange and blue circles show $\mu_\star$-$\Sigma_{\rm SFR}$ for Sa and Sc galaxies, respectively. Only points for which $x_{Y} > 3.4\%$ are included. Diagonal dash-black lines are loci of constant sSFR = 0.01, 0.1, and  1 Gyr$^{-1}$. {\em Right:} the same $\mu_\star$-$\Sigma_{\rm SFR}$ distribution on the left panel but with points weighted with $\log \rho_{\rm SFR}$ (E and S0 are excluded). Lines show the linear fits (for $\alpha^\prime$ and  $\beta^\prime$) to the points in Sa and Sc galaxies, and for all spirals (black-grey). 
}
\label{fig:SFMSlocal}
\end{figure*}

CALIFA is ideally suited to investigate the roles of global and local properties controlling the MSSF. In fact, using the spatially resolved H$\alpha$ flux of more than 500 CALIFA galaxies, Cano-D\'\i az et al.\ (2016) have recently found that the H$\alpha$-based $\Sigma_{\rm SFR}$  correlates with the stellar mass surface density, $\mu_\star$, and that the slope and dispersion of this local MSSF are similar to those of the global SFR-$M_\star$ relation. 
A local MSSF relation  has also been also reported by \citet{wuyts13} in a sample of massive star-forming galaxies at $z \sim 1$ through  spatially resolved H$\alpha$ images provided by HST.}
Section  \ref{sec:GlobalMSSF} presented our version of the  global MSSF (see Fig.\ \ref{fig:SFMSglobal}). Here we use our radial profile data to investigate the local relation.

Fig.\ \ref{fig:SFMSlocal} plots $\Sigma_{\rm SFR}$(R) against $\mu_\star$(R) for the nearly 12 thousand radial bins in our 416 galaxies. The plot is ultimately a collection of 416 $\Sigma_{\rm SFR}(R)$ profiles where the radial coordinate is replaced by $\mu_\star(R)$. Clearly, our \starlight-based $\Sigma_{\rm SFR}(R)$ and $\mu_\star(R)$ values correlate. Dotted diagonals mark lines of constant sSFR. An eye-ball comparison of these lines with the data already hints that, like the global one, the local MSSF is sublinear.

Individual points in the left panel of Fig.\ \ref{fig:SFMSlocal} are color coded according to the galaxy morphology. Large white circles show the mean $\Sigma_{\rm SFR}$ in 0.2 dex wide bins in $\mu_\star$. The scatter around this mean relation is visibly related to morphology, as  further illustrated by the mean relations obtained for Sa and Sc galaxies, drawn as blue and orange circles, respectively. 
The increase in  $\Sigma_{\rm SFR}$ at fixed $\mu_\star$ from early to late types
is another manifestation of our earlier finding that the sSFR$(R)$ profiles scale with Hubble type (Fig.\ \ref{fig:sSFR}).

Galaxies of different morphologies thus seem to follow roughly parallel local MSSF relations of the type $\log \Sigma_{\rm SFR} =  \alpha \log \mu_\star  + \beta$, with similar logarithmic slopes ($\alpha$) but zero points ($\beta$) increasing steadily from early to late types. This behavior prompted us to follow a two-steps approach to estimate $\alpha$ and $\beta$. First, scale effects are removed by rescaling both $\Sigma_{\rm SFR}$ and $\mu_\star$ for each galaxy by their corresponding values at $R = 1$ HLR. The value of $\alpha$ obtained in this way is then used to derive $\beta$ as the average of $\log \Sigma_{\rm SFR} -  \alpha \log \mu_\star$.\footnote{Units of $M_\odot\,{\rm yr} ^{-1}\,{\rm pc} ^{-2}$ and $M_\odot\,{\rm pc} ^{-2}$ are assumed throughout.}

\label{sec:globalMS}
\begin{table}
\caption{Parameters of $\log \Sigma_{\rm SFR}
(M_\odot\,{\rm yr} ^{-1}\,{\rm pc} ^{-2}) = \alpha \log \mu_\star  (M_\odot\,{\rm pc} ^{-2}) + \beta$ fits to the local MSSF to spirals of different types.}
\begin{tabular}{lcccccc}
\hline\hline
 morph. & All & Sa & Sb & Sbc & Sc & Sd    \\      
\hline
$\alpha$               &  $0.70$ & $0.60$  & $0.68$ & $0.70$ & $0.79$  & $0.85$    \\
$\alpha^\prime$   &  $0.84$ & $0.87$  & $0.79$ & $0.78$ & $0.89$  & $0.80$   \\
$\beta$               & $-0.55$ & $-0.70$  & $-0.64$ & $-0.32$ & $-0.41$  & $-0.39$  \\
$\beta^\prime$   & $-0.85$ & $-1.45$  & $-0.93$ & $-0.50$ & $-0.60$  & $-0.30$   \\
 \hline\hline
\label{tab:LocalMSSF}
\end{tabular}
\end{table}

For the whole data set we obtain $\alpha = 0.70 \pm 0.01$ and $\beta = -0.53 \pm 0.02$, with an rms dispersion of  0.27 dex. We have also carried out fits weighting each point by the  $V_{max} ^{-1}$ value of its host galaxy\footnote{In these ``volume-corrected fits'' the weight attributed to all radial bins of the $i^{th}$  galaxy  is  $w_i =  V_{max,i} ^{-1} / \sum_j   V_{max,j} ^{-1}$, where the sum runs over all galaxies.}, which gives the slope and zero point the status of being representative of the local Universe. The parameters for this alternative fit are $\alpha^\prime = 0.84 \pm 0.01$ and $\beta^\prime = -0.85 \pm 0.03$. This fit is shown as a dashed black-grey line in the right panel of Fig.\ \ref{fig:SFMSlocal}, which repeats our local MSSF, but now coloring  each radial bin of each galaxy by its contribution to the total SFR cosmic density in the local Universe ($\rho_{\rm SFR}$).

These fits describe well the local MSSF as a whole, but completely overlook the evident role of morphology. It is thus more appropriate to fit the relation for different Hubble types, in analogy with what was done in Table \ref{tab:GlobalMSSF} for the global MSSF. Table \ref{tab:LocalMSSF} lists the $\alpha$ and $\beta$ values obtained subdividing the sample in morphology. Coefficients for the $V_{max}$-weighted fits ($\alpha^\prime$ and $\beta^\prime$) are also given.  
Dashed blue-grey and orange-grey lines in the right panel of Fig.\ \ref{fig:SFMSlocal} show the fitted relations for Sc and Sa galaxies, respectively.


Inspection of the results in Table \ref{tab:LocalMSSF}  shows that, as anticipated by a visual assessment of the local MSSF, slopes are indeed fairly similar for all types, while $\beta$ increases monotonically from early to late types. 
In all cases we obtain $\alpha < 1$. While the mixture of morphological types certainly explains part of the sub linearity of both the global and local MSSF when lumping all sources together, this result indicates that the local MSSF is sub linear even for fixed Hubble type.
Our values of $\alpha \,(\alpha^\prime) = 0.68 \,(0.79)$ for Sb and 
$0.79 \, (0.89)$ for Sc galaxies bracket the slope of  0.72 derived by Cano-D\'\i az et al.\ (2016) for the H$\alpha$-based local MSSF relation.


Our own previous work has shown that $\mu_\star$ is an effective tracer of local stellar population properties. Both mean stellar ages \citep{gonzalezdelgado14a} and metallicities \citep{gonzalezdelgado14b} correlate well with $\mu_\star$, and this work  shows that 
$\Sigma_{\rm SFR}$ also follows this pattern. 

These previous studies reveal that the overall balance between local ($\mu_\star$-driven) and  global  ($M_\star$-driven) effects varies with the location within the galaxy. While in disks $\mu_\star$ regulates the mean stellar ages and metallicities, it plays a minor role in spheroids (bulges and elliptical galaxies), whose chemical enrichment happened much faster and earlier than in disks. How does the local MSSF relation found in this work fit into this general scheme?

On the one hand, we have seen that the local MSSF relation is mostly a disk phenomenon. In fact, it points to a density dependence of the  SFR law akin to that proposed by \citet{schmidt59} and \citet{kennicutt98}, where  the gas density sets the rate at which stars form. On the other hand, the clear role of Hubble type in defining the offset around the overall $\Sigma_{\rm SFR}$-$\mu_\star$ relation suggests that some global morphology-related property modulates the local sSFR. Gas content is an obvious candidate hidden variable in this context  \citep{roberts94, tacconi13}.
 
Alternatively (or complementarily), the modulation of the $\Sigma_{\rm SFR}$-$\mu_\star$ relation with Hubble type may reflect the effect of a ``morphological quenching''. In \citet{gonzalezdelgado15} we have found that, for the  same $M_\star$, early type galaxies  are older than  later types, both globally and in the disk,  and that this ranking is maintained with radial distance. This gradual age change from spheroidals to Sa and to late spirals reflects the change of  sSFR with Hubble type, and can be interpreted as a consequence of the mechanism building the bulge. The steep potential well induced by the formation of a large spheroid component stabilizes the disk,  cutting the supply of the gas and preventing its local fragmentation into bound, star-forming clumps \citep{martig09}. 
This effect should thus be more significant in  E and S0, and gradually decrease from Sa to Sb. Later types, Sc and Sd, where the bulge (if present) may be formed by secular processes, may not be affected by this  morphological quenching.

\subsection{The relation between local and global MSSFs}
\label{sec:Local2Global}

The results reported throughout this paper give plenty of material to explore in relation with galaxy structure and evolution studies. In this final section we develop some simple math relating the local and global MSSF relations. 

The global (i.e., spatially integrated) SFR and stellar mass of a galaxy relate to the local properties through

\begin{equation}
\label{eq:Model_DefSFR}
{\rm SFR} =  2\pi \int \Sigma_{\rm SFR}(R) R dR = 2\pi R_0^2 \Sigma_{\rm SFR}(R_0)  \,  s_\Sigma
\end{equation}

\begin{equation}
\label{eq:Model_DefM}
M_\star  =   2\pi \int \mu_\star(R) R dR = 2\pi R_0^2 \mu_\star(R_0)  \,  s_\mu
\end{equation}

\noindent where we have denoted HLR by $R_0$ for convenience, and 
\begin{equation}
\label{eq:ModelDef_s_Sigma}
s_\Sigma \equiv  \int \frac{\Sigma_{\rm SFR}(R)}{ \Sigma_{\rm SFR}(R_0) }  \frac{R}{R_0} \frac{dR}{R_0}
\end{equation}

\begin{equation}
\label{eq:ModelDef_s_mu}
s_\mu \equiv  \int \frac{ \mu_\star(R) }{ \mu_\star(R_0) }  \frac{R}{R_0} \frac{dR}{R_0}
\end{equation}

\noindent are shape factors of order unity. Equations  \ref{eq:Model_DefSFR} and  \ref{eq:Model_DefM} 
lead to 

\begin{equation}
\label{eq:Model_SFR}
{\rm SFR} =  
\frac{ s_\Sigma }{ s_\mu } \frac{ \Sigma_{\rm SFR}(R_0) }{ \mu_\star(R_0)} M_\star 
\end{equation}

\noindent which predicts the global MSSF relation in terms of spatially resolved properties.\footnote{Eq.\ \ref{eq:Model_SFR} can be written more compactly as ${\rm sSFR} = \frac{ s_\Sigma }{ s_\mu } {\rm sSFR}(R_0)$, where the left-hand-side is the global (spatially integrated) sSFR.}

Direct integration of the profiles yields a $s_\Sigma / s_\mu$ ratio of typically $0.9 \pm 0.4$ for our spirals (average and dispersion), and a very weak
($\propto M_\star^{-0.07}$)
 trend with mass. Relevant deviations from a linear global MSSF must therefore come from variations of ${\rm sSFR}(R_0) = \Sigma_{\rm SFR}(R_0) / \mu_\star(R_0)$ with $M_\star$.

Fig.\ \ref{fig:sSFR} shows that ${\rm sSFR}(R_0)$ increases systematically towards later type spirals, indicating an anti correlation with stellar mass and hence a sub-linear predicted global MSSF. 
More quantitatively, recalling that  $\Sigma_{\rm SFR} \propto \mu_\star^\alpha$ from our local MSSF relation, and that $\mu_\star(R_0) \propto M_\star^\gamma$, with $\gamma \sim 0.5$ \citep{gonzalezdelgado14a}, the predicted relation goes as ${\rm SFR} \propto M_\star^{1 -  \gamma (1 - \alpha)}$. For $\alpha$  between 0.70 and 0.84 (Table \ref{tab:LocalMSSF}), and correcting for the mild trend of $s_\Sigma / s_\mu$ with mass, the predicted logarithmic slope of the global MSSF is in the 0.78--0.85 range, in good agreement with \citet{renzinipeng15} and Cano-D\'\i az et al.\ (2016).

We close by noting that it is plausible to conclude from this analysis that the sub-linearity of the local MSSF ($\alpha < 1$) is what causes sub-linearity of the global MSSF\footnote{
We clarify that the sub-linearity we refer to here is not that resulting from mixing galaxies of different morphologies in a same sample, but 
the one found when fitting the global MSSF {\em at fixed Hubble type} (i.e., the $a < 1$ slopes in Table \ref{tab:GlobalMSSF}). 
}.
A caveat in this tempting local $\rightarrow$ global argument is that 
it
uses $\mu_\star$ to trace the local SFR density, whereas gas, not stars, is the actual fuel of star formation. We thus postpone further analysis of this issue to future work involving gas density estimates.

\section{Summary and conclusions}
\label{sec:Summary}

We analyzed the stellar population properties of 416 galaxies, observed by CALIFA at the 3.5m telescope in Calar Alto, to investigate the trends of the recent star formation rate with radial distance and as a function of  Hubble type. The sample includes ellipticals, S0, and spirals all the way from Sa to Sd, covering a stellar mass range from 
$\sim10^9$ to $7\times10^{11} M_\odot$ (for a Salpeter IMF). A full spectral fitting analysis was performed using the \starlight\ code and a combination of SSP spectra from \citet{gonzalezdelgado05} plus \citet{vazdekis10}. Our \pycasso\ pipeline was used to process the spectral fitting results to produce maps of the recent star formation rate (SFR, averaged over the last 32 Myr), and the stellar mass surface density ($\mu_\star$).  For each galaxy, the maps are azimuthally averaged to produce radial profiles (in units of the half light radius, HLR) of the SFR surface density, $\Sigma_{\rm SFR}(R)$, and the corresponding local specific SFR, sSFR$(R) = \Sigma_{\rm SFR}(R)/\mu_\star(R)$.  Variations of the traditional birthrate  parameter, $b$, are obtained to compare the present and the past SFR at different radial positions. The radial profiles are stacked as a function of Hubble type and of galaxy mass to identify the main trends.

Our main results are:

\begin{enumerate}

\item
Spiral galaxies have declining $\Sigma_{\rm SFR}(R)$ profiles, with a relatively tight range of $\Sigma_{\rm SFR}$ values at any given radial distance. At $R =  1$ HLR the $\Sigma_{\rm SFR}$ is  typically $20\, M_\odot\,$Gyr$^{-1}\,$pc$^{-2}$, with a factor of two dispersion.
Spirals with  $M_\star \la  4 \times 10^{10} M_\odot$  have $\Sigma_{\rm SFR}(R)$ profiles that are very similar and independent of  Hubble type and galaxy mass. Above $4 \times10^{10} M_\odot$ the $\Sigma_{\rm SFR}(R)$ profiles are slightly more dispersed.  
This is a remarkable result taking into consideration that the sample covers two orders of magnitude in $M_\star$ and all Hubble types.
Ultimately, it is the constancy of $\Sigma_{\rm SFR}$ that, coupled to the $\mu_\star$-$M_\star$ relation, makes the MSSF a tight sequence.

\item 
In contrast, E and S0 galaxies have $\Sigma_{\rm SFR}(R)$  that are at all radii significantly depressed with respect to spirals, with flat $\Sigma_{\rm SFR}$ $\sim1 M_\odot\,$Gyr$^{-1}\,$pc$^{-2}$  profiles,  with a large uncertainty. 

\item 
Expressed in units of the lifetime averaged SFR intensity at each location, the present $\Sigma_{\rm SFR}(R)$ is currently lower in E, S0, and early type spirals (Sa and Sb), but higher in later spirals (Sc and Sd). 
Sbc galaxies seem to be the transition type in which "bulges" (central $\sim$1 HLR) have already suppressed/quenched their star formation activity, as in Sa and Sb, but their disks are still forming new stars at a rate similar to the past. 

\item The local ${\rm sSFR} = \Sigma_{\rm SFR}/\mu_\star$ shows radial profiles that  increase outwards and scale with Hubble type, from Sa to Sd. This behavior is preserved at any given $R$. This quantity, that relates locally the present and the past star formation rate, is orders of magnitude smaller in E and S0 than in spirals.
The characteristic time scale of star formation given by sSFR$^{-1}$ in spirals ranges from 12.6 Gyr in Sbc to 5 Gry in Sd galaxies. Early type spirals (Sa, Sb) and spheroidals (E, S0) would need more than a Hubble time to build  their current stellar mass at their recent SFR.  

\item The slope of sSFR$(R)$ in the inner 1 HLR is steeper than outwards. This behavior with radial distance suggests that  galaxies are quenched inside-out, and that this process is faster in the central part (dominated by the bulge) that in the disk. 
 
\item
The CALIFA sample is well suited to compute the SFR density in the local universe, with a value  $\rho_{\rm SFR} = 0.0105\pm0.0008$ (random) $M_\odot\,$yr$^{-1}\,$Mpc$^{-3}$ (for a Salpeter IMF),  in excellent agreement with previous estimates from completely difference methods and data.
We find the majority of the star formation at $z=0$ to take place in Sbc, Sc, and Sd galaxies with masses below $10^{11}  M_\odot$. In terms of spatial distribution, most of the star formation is occurring outside galaxy centres, in regions that are mainly in the disks of spirals.

\item
The volume average birthrate parameter, $b^\prime = 0.39$, suggests that the present day Universe is forming stars at $\sim 1/3$ of its past average rate.  E, S0, and the bulge of early type spirals have $b < 0.39$, thus  contributing little to the present star formation rate of the Universe. The disks (regions outside 1 HLR) of Sbc, and Sc, and Sd galaxies, all with $b >1$, dominate the present star formation of the Universe.

\item
Galaxy mass and morphology, in particular the formation of a spheroidal component, play a relevant role in depressing/quenching the star formation in galaxies. Galaxies dominated by the spheroidal component, E and S0 in our sample, are all quiescent. Disk dominated galaxies (Sbc, Sc, Sd) are very actively forming stars with a rate per unit mass that decreases with $M_\star$. 
 
\item
There is tight relation between the local values of $\mu_\star$ and $\Sigma_{\rm SFR}$, defining a local main sequence of star forming regions with slope $\sim 0.8$, and a scatter strongly related to Hubble type. 
This relation is tighter than the global main sequence relation between SFR and $M_\star$
 
once morphology-related offsets are accounted for.
This suggests that local processes are important in determining the star formation in a galaxy, possibly due to a density dependence of the SFR law.  
The shut down of the star formation is more related with global processes, such as the formation of a spheroidal component. These findings agree with our previous analysis that showed that the mean stellar ages and metallicity are mainly governed by local processes ($\mu_\star$-driven) in disks, and by global processes in spheroids. 

\end{enumerate}

Thanks to the uniqueness of CALIFA data and  the homogeneity of our analysis,  we were able to, for the first time, characterize the radial structure of the star formation rate along the Hubble sequence. 
This octogenarian sequence, by the way, has once again demonstrated its usefulness as a way to organize galaxies in terms of their spatially resolved properties. Our previous work showed the systematic behavior of $\mu_\star$, mean stellar ages and metallicity with Hubble type, while this paper showed the same also applies to the SFR and related quantities.

The large file FoV of PPaK allows us to cover galaxies in their entire optical extent, a design feature of CALIFA which eliminates aperture-related biases in the derivation of galaxy properties. Furthermore, the well defined selection function of the survey allows for reliable volume corrections. In fact, an important ``byproduct" of this study is that it exemplifies how well these corrections work, as demonstrated by the excellent agreement between our estimate for the $\rho_{\rm SFR}$ of the local Universe and independent determinations from large galaxy surveys. This validates and reinforces the statistical approach to CALIFA followed in this, previous and future papers in this series.

\begin{acknowledgements} 
CALIFA is the first legacy survey carried out at Calar Alto. The CALIFA collaboration would like to thank the IAA-CSIC and MPIA-MPG as major partners of the observatory, and CAHA itself, for the unique access to telescope time and support in manpower and infrastructures.  We also thank the CAHA staff for the dedication to this project.
Support from the Spanish Ministerio de Econom\'\i a y Competitividad, through projects AYA2014-57490-P, AYA2010-15081, and Junta de Andaluc\'\i a FQ1580, AYA2010-22111-C03-03, AYA2010-10904E, AYA2013-42227P, RyC-2011-09461, AYA2013-47742-C4-3-P, EU SELGIFS exchange programme FP7-PEOPLE-2013-IRSES-612701, and CONACYT-125180 and DGAPA-IA100815.
We also thank the Viabilidad, Dise\~no, Acceso y Mejora funding program, ICTS-2009-10, for funding the data acquisition of this project. 
ALdA, EADL and RCF thanks the hospitality of the IAA and the support of CAPES and CNPq. RGD acknowledges the support of CNPq (Brazil) through Programa Ci\^encia sem Fronteiras (401452/2012-3). CJW acknowledges support through the Marie Curie Career Integration Grant 303912. We thank the support of the IAA Computing group, and to the referee for useful comments.
\end{acknowledgements}



\bibliographystyle{aa}
\bibliography{Califa6_SFR}

\appendix
\section{Dependence of  SFR on SSP models}

\label{sec:SSP}

\begin{figure*}
\includegraphics[width=\textwidth]{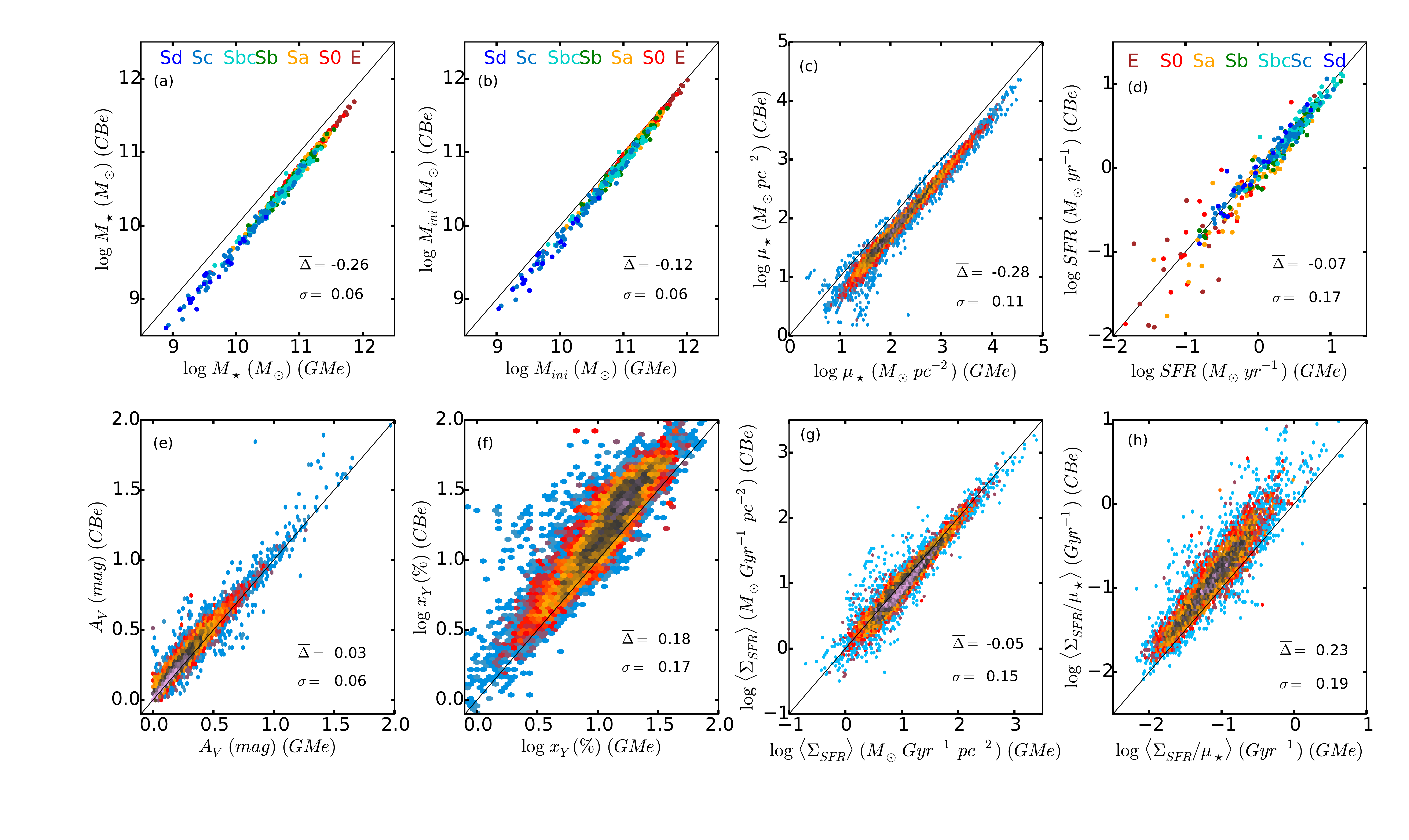}
\caption{
Comparison of several stellar population properties as obtained with the bases {\em GMe} (x-axis) and {\em CBe} (y-axis). The average difference between the property in the y and x-axis is labeled as $\overline{\Delta}$ in each panel, and its standard deviation as $\sigma$.  Panels (a), (b), and (d) show the galaxy mass and SFR, with galaxies colored by their Hubble type. In the other panels, the values of the property measured  every 0.1 HLR are compared, and the color indicates the density of points in a logarithmic scale.
 } 

\label{fig:SFR_bases}
\end{figure*}

In order to evaluate to what extent our results depend on the choice of SSP models, we compare the properties derived with two bases: Base {\em GMe}, i.e., the one used in the main text and briefly described in Sec.\ \ref{sec:Base}, and base {\em CBe}, used in several earlier works by our group, and fully described in \citet{gonzalezdelgado15}. In short, this base is built out of a preliminary update of the \citet{bruzual03} models (Bruzual 2007, private communication), from which we draw $N_\star$ = 246 elements with 41 ages (from 0.001 to 14 Gyr) and six metallicities  ($\log Z/Z_\odot = -2.3$, $-1.7$, $-0.7$, $-0.4$, $0$, and $+0.4$). The evolutionary tracks are those collectively referred to as Padova 1994 by  \citet{bruzual03}, and the IMF is that of \citet{chabrier03}. Compared to {\em GMe}, base {\em CBe} differs in evolutionary tracks, IMF, and metallicity range. 
 
We make two types of comparisons in this appendix: (i) global (galaxy wide) quantities, such as the present day and initial stellar masses, and the total SFR; and (ii) radial averages of $\mu_\star$, $A_V$, $\Sigma_{\rm SFR}$,  $\Sigma_{\rm SFR}/\mu_\star$, and $x_{Y}$ for up to a maximum 30 of points for each galaxy (corresponding to $R = 0$--3 in steps of 0.1 HLR). Fig. \ref{fig:SFR_bases} shows the results, with base {\em GMe} values in the x-axis and  {\em CBe} ones in the y-axis. Each panel shows a one-to-one line, as well as the mean ($\overline{\Delta}$) and standard deviation ($\sigma$) of the difference $\Delta \equiv $ property(CBe) $-$ property(GMe).

On average, {\em GMe}-based $M_\star$ and $\mu_\star$-values are $\sim$0.26 dex higher than the corresponding {\em CBe}-based values, reflecting the different IMF used. Discounting this offset, the two values of stellar mass and mass surface density agree to within 0.06 and 0.11 dex, respectively. In terms of the initial mass that is converted into stars ($M_{ini}$) there is a difference of 0.12 dex between the two bases and a dispersion of  0.06 dex. Again, the difference reflects the change of IMF between the two bases. Note that $\Delta \log M_\star$ is higher than $\Delta \log M_{ini}$ because the returned fraction ${\cal R}$ also differs from one base to the other (${\cal R}$= 0.28 and 0.48 for  {\em GMe} and  {\em CBe}, respectively). 

Due to the IMF difference, the SFR should be lower for {\em CBe} than for {\em GMe}. This is in fact the result (Fig.\ref{fig:SFR_bases}d), but the difference is only 0.07 dex, lower than what we would expect due to the change of IMF. This implies that besides the IMF, there are differences in the SFH between {\em GMe} and {\em CBe} and/or in  stellar extinction. The latter explanation does not hold, since $A_V$ is very similar in the two sets of models, with an offset of only $\Delta = 0.03$ mag and dispersion $\sigma = 0.06$ mag. However, we note that there is an important difference between the light fraction in populations younger than 32 Myr. On average, $x_{Y}$ is  0.18 dex higher with {\em CBe} than {\em GMe}. This explains why the SFR with {\em CBe}, although lower than with {\em GMe}, is not a full factor of $\sim 1.7$ lower, as expected due to the change of IMF. 

This change in SFH, and in particular in $x_{Y}$, does not produce any significant effect in the radial distribution of the star formation rate intensity, $\Sigma_{\rm SFR}(R)$. The two sets of values are well correlated (Fig.\ \ref{fig:SFR_bases}g), with a tiny difference of $\Delta = -0.05$ dex (lower in {\em CBe} than in {\em GMe}), and a dispersion $\sigma =0.15$ dex. The offset of $\Sigma_{\rm SFR}/\mu_\star$  between the two bases is $\Delta = 0.23$ dex, reflecting mainly the offset in $\mu_\star$ due to the IMF (Fig.\ \ref{fig:SFR_bases}b).

\end{document}